\documentclass[
superscriptaddress,reprint,amsmath,amssymb,aip,pra,floatfix]{revtex4-1}
\usepackage[margin=1.5cm]{geometry}

\usepackage{xcolor} 
\usepackage{graphicx}
\usepackage{dcolumn}
\usepackage{upgreek}
\usepackage{bm}
\usepackage{hyperref}
\usepackage{floatrow}
\usepackage{times}
\usepackage[mathlines]{lineno}
\usepackage{physics}
\usepackage{cancel} 
\usepackage{soul}

\usepackage{etoolbox}
\makeatletter
\newcommand*{\newbibstartnumber}[1]{%
  \apptocmd{\thebibliography}{%
    \global\c@NAT@ctr #1\relax
    \addtocounter{NAT@ctr}{-1}%
  }{}{}%
}
\makeatother

\begin{document}
\long\def\/*#1*/{}

\title{Protecting a Bosonic Qubit with Autonomous Quantum Error Correction}
\author{Jeffrey M.~Gertler}
\affiliation{Department of Physics, University of Massachusetts-Amherst, Amherst, MA, 01003, USA}
\author{Brian Baker}
\affiliation{Department of Physics and Astronomy, Northwestern University, Evanston, IL, 60208, USA}
\author{Juliang Li}
\affiliation{Department of Physics, University of Massachusetts-Amherst, Amherst, MA, 01003, USA}
\author{Shruti Shirol}
\affiliation{Department of Physics, University of Massachusetts-Amherst, Amherst, MA, 01003, USA}
\author{Jens Koch}
\affiliation{Department of Physics and Astronomy, Northwestern University, Evanston, IL, 60208, USA}
\author{Chen Wang*}
\affiliation{Department of Physics, University of Massachusetts-Amherst, Amherst, MA, 01003, USA}
\email{wangc@umass.edu}

\date{\today}


\begin{abstract}
To build a universal quantum computer from fragile physical qubits, effective implementation of quantum error correction (QEC)~\cite{lidar_quantum_2013} is an essential requirement and a central challenge. Existing demonstrations of QEC are based on an active schedule of error syndrome measurements and adaptive recovery operations~\cite{schindler_experimental_2011, cramer_repeated_2016, 
kelly_state_2015, ofek_extending_2016, hu_quantum_2019, andersen_repeated_2020
} that are hardware intensive and prone to introducing and propagating errors. 
In principle, QEC can be realized autonomously and continuously by tailoring dissipation within the quantum system~\cite{lidar_quantum_2013, ahn_continuous_2002,atalaya_continuous_2020,kerckhoff_designing_2010,kapit_hardware-efficient_2016,reiter_dissipative_2017, albert_pair-cat_2019, sarovar_continuous_2005}, but so far it has remained challenging to achieve the specific form of dissipation to counter the most prominent errors in a physical platform.
Here we encode a logical qubit in Schr{\"o}dinger cat-like multiphoton states~\cite{brune_observing_1996} of a superconducting cavity, and demonstrate a corrective dissipation process that stabilizes an error syndrome operator: the photon number parity.  Implemented with continuous-wave control fields only, this passive protocol realizes autonomous correction against single-photon loss and boosts the coherence time of the multiphoton qubit by over a factor of two.  Notably, QEC is realized in a modest hardware setup with neither high-fidelity readout nor fast digital feedback, in contrast to the technological sophistication required for prior QEC demonstrations.  Compatible with additional phase-stabilization and fault-tolerant techniques~\cite{mundhada_experimental_2019, reinhold_error-corrected_2020, ma_error-transparent_2020}, our experiment suggests reservoir engineering as a resource-efficient alternative or supplement to active QEC in future quantum computing architectures.  
\end{abstract}

\maketitle

Robustness in modern classical computers is often aided by passive dissipation acting as a restoring force against environmental perturbations, rather than active error correction protocols. 
A quantum analogy for dissipative stabilization of qubits, or autonomous quantum error correction (AQEC), is possible but has important distinctions from its classical counterpart (Fig.~1a): Since qubit states form a continuous two-dimensional encoding manifold $\mathcal{C}$,  the restoring dissipation must be tailored perpendicular to $\mathcal{C}$ to not disrupt the manifold of code states, which leaves any perturbations tangential to $\mathcal{C}$ unrecoverable. Therefore, the encoding manifold $\mathcal{C}$ itself also has to be specifically designed to ensure that the physical errors only occur orthogonal to the surface, as formulated by the Knill-Laflamme QEC criteria~\cite{knill_theory_1997}.

\begin{figure}[tbp]
    \centering
    \includegraphics[scale=0.54]{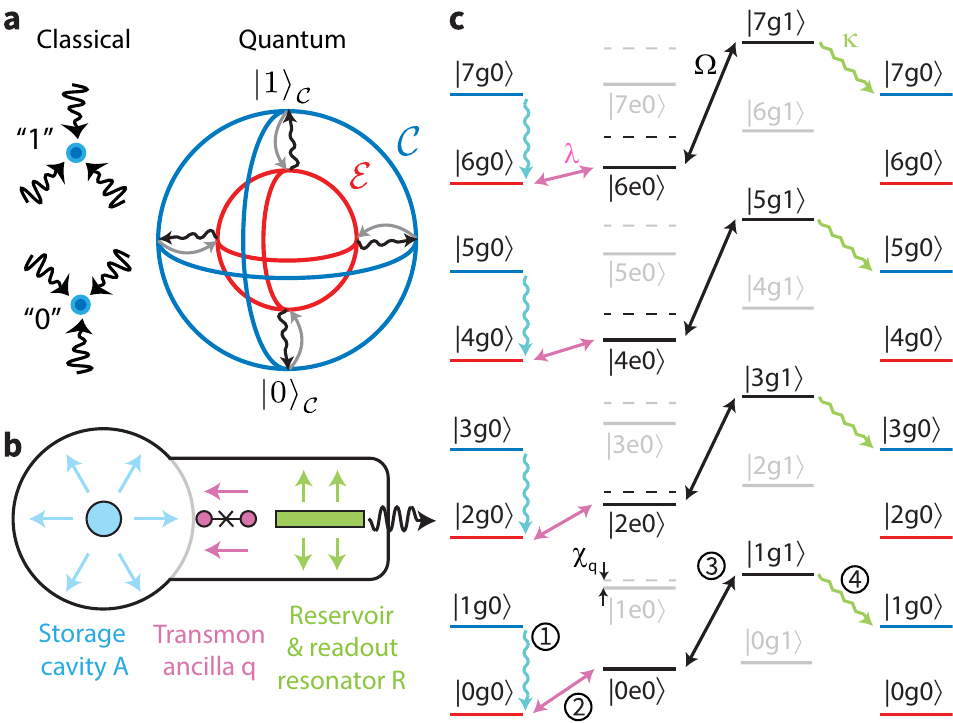}
    \caption{\textbf{Autonomous quantum error correction: concept and protocol.}  \textbf{a}, A classical bit is stored in isolated points in configuration space, which are local energy minima stabilized by dissipation in all directions.  In comparison, a logical qubit $\ket{\psi}$ is encoded in a continuous two-dimensional code space $\mathcal{C}$ designed in a way that both natural errors (which map the states to the error space $\mathcal{E}$) and engineered dissipation are only allowed perpendicular to $\mathcal{C}$.
    \textbf{b}, Schematic of the circuit QED device composed of storage cavity $A$, transmon ancilla $q$ and reservoir resonator $R$.
    \textbf{c}, AQEC scheme against single-photon loss illustrated in a level diagram (not to scale). The level indices refer to $A$, $q$, $R$ sequentially.  A continuous wave (cw) ``transmon comb" is applied to resonantly excite the transmon ancilla with a Rabi rate $\lambda$ (magenta arrows), selectively targeting the four even-parity states (red levels) when $\chi_q \gg \lambda$.  Similarly, a cw ``mixing comb" targets the $\ket{2n,e,0}\leftrightarrow\ket{2n+1,g,1}$ transitions with an equal Rabi rate $\Omega$ (black arrows).  Both combs are composed of four tones equally spaced by $2\chi_q$ as indicated by the varying slopes of the magenta and black transitions.  Spontaneous decay of the reservoir $R$ converts the quantum state back to the code space (blue levels) without leakage of which-path information; note the identical slopes of the light-green arrows.  The numbered labels show the error and recovery sequence \textcircled{\raisebox{-0.8pt}{1}}-\textcircled{\raisebox{-0.8pt}{4}} for one of the four parallel paths (using the initial state $\ket{1g0}$ as an example). 
}
\end{figure}

These simultaneous requirements dictate that the form of dissipation needed for AQEC will be rather exotic~\cite{lihm_implementation-independent_2018}.  Advances in quantum reservoir engineering~\cite{poyatos_quantum_1996} have paved the way to synthesize dissipation operators not naturally available,  
enabling stabilization of various non-classical states~\cite{krauter_entanglement_2011, barreiro_open-system_2011, kienzler_quantum_2015}.  More recently, dissipation has been tailored to confine a quantum harmonic oscillator to delocalized regions in phase space~\cite{leghtas_confining_2015} to suppress bit-flips~\cite{lescanne_exponential_2020}. 
However, the stabilized manifold does not permit QEC of the dominant decoherence process in the system: single photon loss. 
Encouraging proposals for AQEC have emerged for a number of experimental platforms~\cite{kerckhoff_designing_2010,kapit_hardware-efficient_2016,reiter_dissipative_2017,albert_pair-cat_2019}, but they require hardware architectures yet to be developed.  In this Article, we introduce and implement an AQEC scheme in a common transmon-based superconducting circuit QED device~\cite{koch_charge-insensitive_2007, reagor_quantum_2016, axline_architecture_2016}, 
which is solely enabled by the synthesis of a highly-specific dissipation operator. 

\subsection*{Error correction code and strategy}
Following the accelerating progress in bosonic QEC research~\cite{chuang_bosonic_1997, terhal_towards_2020}, we take advantage of the large Hilbert space and the long coherence time~\cite{reagor_quantum_2016} of microwave-photon states in a superconducting cavity to store a logical qubit.
The qubit, $\ket{\psi_L}=x\ket{0_L}+y\ket{1_L}$ (with $x$ and $y$ being complex coefficients containing the logical information), is encoded in an odd-parity subspace of the cavity 
using the following Truncated 4-component Cat  (T4C) code~\cite{mirrahimi_dynamically_2014}:
\begin{align}
    \ket{0_L}=C_1\ket{1}+C_5\ket{5},\,\,\,\,\,
    \ket{1_L}=C_3\ket{3}+C_7\ket{7},
    \label{eq:code}
\end{align}
where $C_1,C_3,C_5,C_7=\sqrt{0.35},\sqrt{0.9},\sqrt{0.65},\sqrt{0.1}$ are code word coefficients of the four Fock-state components, chosen to approximately balance the average photon numbers in $\ket{0_L}$ and $\ket{1_L}$ ($\bar{n}\approx3.5$). 
The $\hat{x}$ basis states (instead of $\hat{z}$) of the code, $\ket{\pm_L}=(\ket{0_L}\pm\ket{1_L})/\sqrt{2}$, resemble the Schr{\"o}dinger cat superpositions of coherent states, $\ket{C_\alpha}=(\ket{\alpha}-\ket{-\alpha})/\mathcal{N}$.
The coherent-state amplitude, $|\alpha|\approx1.87$, is a measure of the size of the cat-state encoding.  Single-photon loss, the dominant intrinsic error in a superconducting cavity to be corrected, converts odd-numbered Fock components to even-numbered ones.  The photon-number parity $\hat{P}=e^{i \pi \hat{a}^\dagger \hat{a}}$ effectively plays the role of a QEC stabilizer operator, which can be repetitively measured in active QEC protocols~\cite{ofek_extending_2016, hu_quantum_2019}. 

\begin{figure*}[bt]
    \centering
    \includegraphics[scale=0.50]{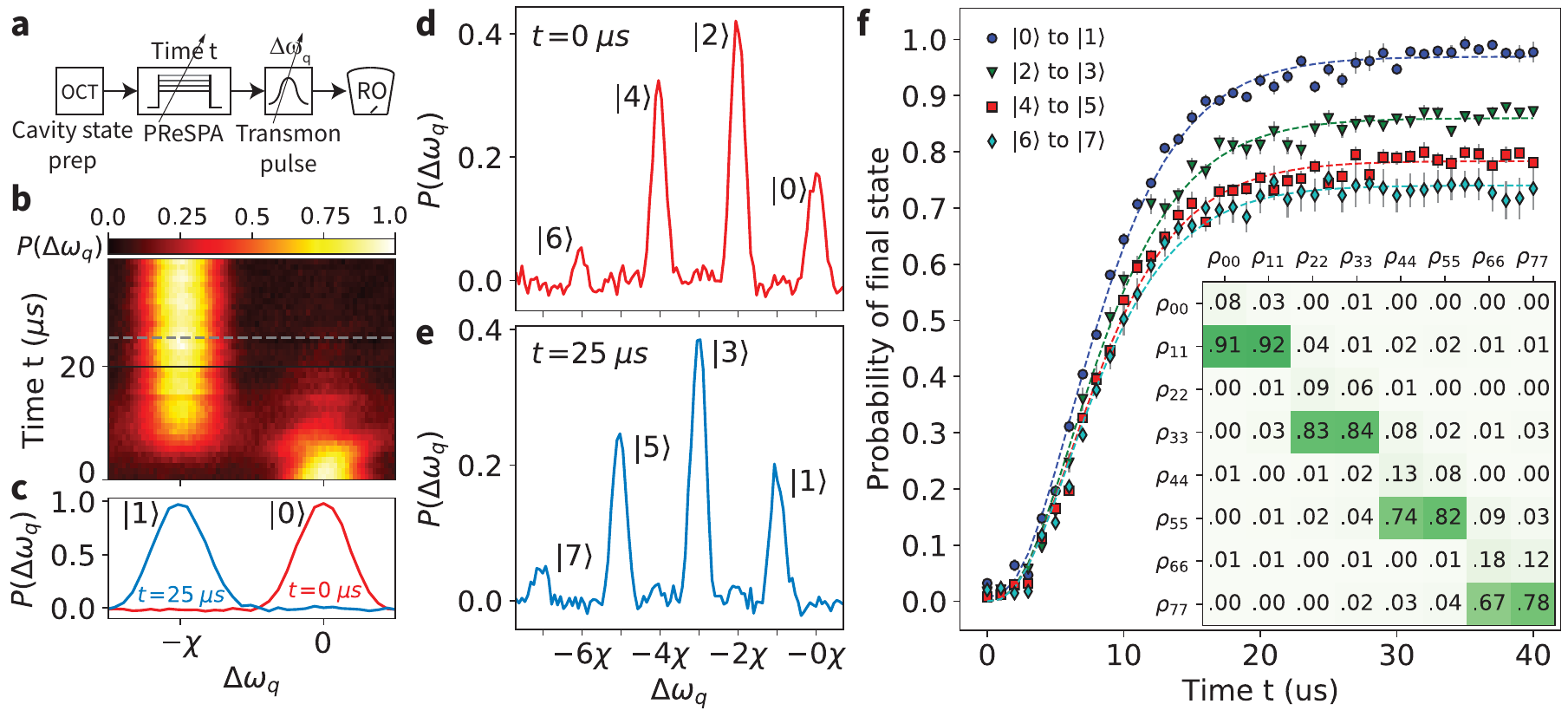}
    \caption{\textbf{Characterization of the PReSPA operator: photon population conversion.}  \textbf{a}, Control pulse sequence of a transmon spectroscopy measurement to infer the cavity photon distribution after PReSPA: We initialize cavity $A$ to a specific initial state using an Optimal Control Theory (OCT) pulse~\cite{heeres_implementing_2017}, apply PReSPA for a variable time $t$, apply a spectrally-selective $\pi$-pulse to transmon $q$ at a variable detuning $\Delta \omega_q$, and measure the transmon excitation probability $P(\Delta\omega_q, t)$. 
    \textbf{b}, Transmon spectroscopy data $P(\Delta\omega_q, t)$ for cavity $A$ initialized in vacuum.  The bright feature is shifted from $\Delta\omega_q=0$ to $-\chi_q$ over time, showing the $\ket{0}_A\rightarrow\ket{1}_A$ conversion.  (For $0<t<20$\,$\mu$s an additional delay time of 20 $\mu$s is inserted between PReSPA pumps and the transmon $\pi$-pulse to improve clarity of the spectroscopy data by allowing the partially-excited transmon to relax.)  
    \textbf{c}, Cuts of Fig.~2b at $t=0$ and $25$ $\mu$s (grey dashed line). 
    \textbf{d}, \textbf{e}, $P(\Delta\omega_q)$ for $A$ initialized in an even-parity cat state at \textbf{d}, $t=0$ and \textbf{e}, $t=25$ $\mu$s.  All four spectroscopy peaks corresponding to even photon numbers (red) are shifted by $-\chi_q$ after $t=25$ $\mu$s indicating odd photon numbers (blue). 
    \textbf{f}, Probability of achieving the target cavity state $\ket{2n+1}_A$, as measured by $P(t)$ for fixed $\Delta\omega_q=-(2n+1)\chi$, for cavity $A$ initialized in $\ket{2n}_A$.  Error bars reflect standard error of the mean.  These four time-domain curves are fitted using a numerical model of the cascaded pumping process (see Methods and Extended Data Fig.~3), resulting in $\Omega$ = 92, 88, 87, 85 kHz; and $\lambda$ = 28, 27, 27, 26 kHz, respectively. The inset shows a block of the cavity process $\chi$ matrix for 25\,$\mu$s of PReSPA.  The matrix elements $\chi_{nn,n'n'}$ are calculated from transmon spectroscopy measurements from all pairs of initial Fock states $\ket{n}_A$ and final Fock states $\ket{n'}_A$.}
\end{figure*}

The main technical accomplishment of this work is the realization of \textit{Parity Recovery by Selective Photon Addition} (PReSPA) via a constantly-applied dissipation operator
\begin{equation}
\hat{\Pi}_{eo}=
\ket{1}\bra{0}+\ket{3}\bra{2}+\ket{5}\bra{4}+\ket{7}\bra{6}.
\label{eq:PReSPA}
\end{equation}
This operator stabilizes the photon-number parity while preserving coherence between the Fock components.  Whenever a parity jump arises from photon loss, PReSPA performs AQEC by automatically adding a photon back to the cavity. 
PReSPA is not constructed from a true parity operator ($\propto e^{i \pi \hat{a}^\dagger \hat{a}}$). However, by explicitly constructing a superposition of four targeted dissipative processes, it achieves a workaround to the well-recognized challenge of implementing continuous quantum non-demolition parity projection~\cite{albert_pair-cat_2019,cohen_degeneracy-preserving_2017}, which has been a major obstacle for AQEC.  
It should be noted that $\hat{\Pi}_{eo}$ does not fully reverse the effect of a photon loss event ($\hat{\Pi}_{eo}\hat{a}\neq \hat{\mathrm{I}}$); rather it leads to a net gain in mean photon numbers for all logical states, which is approximately equivalent to an increase of
$|\alpha|$. PReSPA also does not correct for the continuous decrease of $|\alpha|$ in the absence of parity jumps~\cite{ofek_extending_2016}.  
However, the accumulated entropy in the size uncertainty of the cat-state encoding is not associated with significant leakage of logical information (which is encoded in the phases of the cat states), and can be 
removed from the logical qubit effectively using a proper decoding transformation.  Theoretically, instantaneous and exact $\hat{\Pi}_{eo}$ operations to this T4C code can reduce its logical error rate from single-photon loss by $\sim$50 times (or more if a larger $|\alpha|$ is chosen for the encoding).  See Methods for the theory of this approximate AQEC protocol.

\subsection*{AQEC technique and device implementation}
Our experiment is carried out in a 3D-planar hybrid circuit QED architecture~\cite{axline_architecture_2016} at a base temperature of about 10 mK.  A high-coherence cylindrical post cavity $A$ (with $T_{1A}=520$ $\mu$s, $T_{2A}=380$ $\mu$s) is used to store the logical qubit~\cite{reagor_quantum_2016}.  A dispersively-coupled transmon~\cite{koch_charge-insensitive_2007} qubit $q$ (with $T_{1q}=39$ $\mu$s, $T_{2q}^*=17$ $\mu$s) is used as an ancilla for encoding and decoding of the cavity state.  A coaxial stripline resonator $R$ with fast decay rate ($\kappa/2\pi=0.58$ MHz) is used both for readout and as the source of dissipation (Fig.~1b).  The leading-order terms of the system Hamiltonian are: 
\begin{align}
    \frac{\hat{H}}{\hbar} = &\omega_A \hat{a}^\dagger \hat{a} + \frac{\omega_q}{2}\hat{\sigma}_z  
    + \omega_R \hat{r}^\dagger\hat{r} - \frac{\chi_q}{2} \hat{a}^\dagger\hat{a} \hat{\sigma}_z  
    -\frac{\chi_r}{2}\hat{r}^\dagger \hat{r}\hat{\sigma}_z ,
    \label{eq:staticH}
\end{align} 
where $\hat{a}$ and $\hat{r}$ are the lowering operators in $A$ and $R$, $\sigma_z$ is the Pauli operator of the transmon. The dispersive shift between modes $q$ and $A$ is $\chi_q/2\pi=1.32$ MHz and between modes $q$ and $R$ is $\chi_r/2\pi=2.8$ MHz. 
The PReSPA operator is implemented with a four-fold degenerate two-stage pumping process, as illustrated in Fig.~1c.  
Two continuous-wave (cw) frequency combs, each consisting of four tones equally spaced in frequency by $2\chi_q$, are applied to drive transitions targeting the four even-number Fock states, $\ket{2n}_A$ ($n=0,1,2,3$).  Under the rotating wave approximation, the drive Hamiltonian is
\begin{align}
 \frac{\hat{H}_d}{\hbar} = 
    \sum_{n=0}^3 &\big( \lambda\ket{2n,e,0}\bra{2n,g,0} \nonumber\\
    &+ \Omega\ket{2n+1,g,1}\bra{2n,e,0}\big) + \text{h.c.}
    \label{eq:drivenH}
\end{align}
The selectivity on individual levels relies on the photon-number dependent transition frequencies of the dispersive Hamiltonian. Hence, it is critical that 
$\lambda\ll\chi_q$, so the odd-parity code space remains unperturbed.  The other rates follow a hierarchy of $\lambda<\Omega<\kappa<\chi_q$.
When a photon emission occurs in $R$, the even-parity states are simultaneously projected to the odd-parity subspace by gaining a photon.  By adiabatic elimination of the fast dynamics in $q$ and $R$, 
we obtain $\hat{\Pi}_{eo}$ as the effective operator acting on cavity $A$.  Phase coherence among the four converted states is expected to persist,  
since no which-path information leaks into the environment as long as 1) the transition rates along the four paths are identical, and 2) the frequency of the emitted reservoir photon is independent of the path choice. 
This path-degenerate pumping process is inspired by several theoretical proposals~\cite{sarovar_continuous_2005, pinotsi_single_2008, kapit_hardware-efficient_2016}, but our two-stage construction of the coherent drives is crucial for achieving path-independent reservoir emission frequencies while maintaining parity selectivity. 

\begin{figure*}[tbp]
    \centering
    \includegraphics[scale=.53]{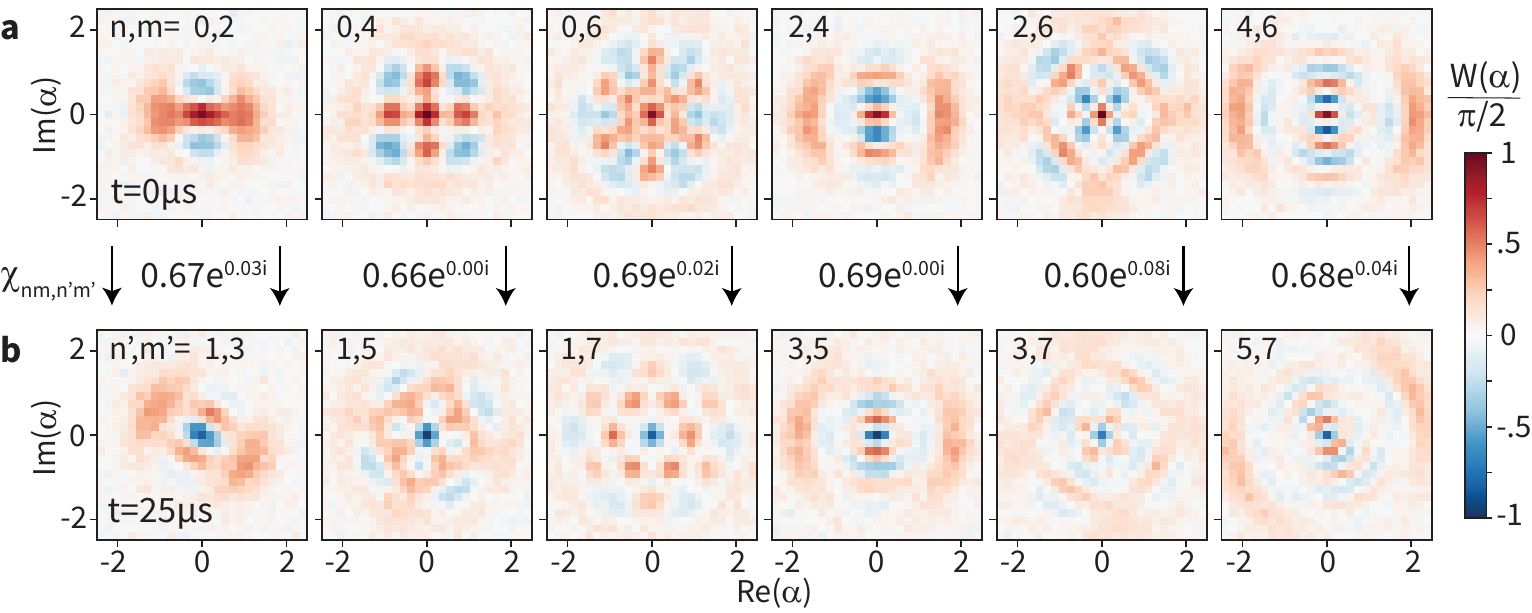}
    \caption{\textbf{Characterization of PReSPA operator: preservation of coherence.}  \textbf{a}, Cavity Wigner tomography of six even-parity superposition states, $(\ket*{n}_A+\ket*{m}_A)/\sqrt{2}$, prepared by OCT pulses, as input states for PReSPA.  \textbf{b}, Wigner tomography of the six corresponding output states after 25 $\mu$s of PReSPA, which are converted approximately to odd-parity superpositions $(\ket*{n'}_A+\ket*{m'}_A)/\sqrt{2}$, with $n'=n+1$, $m'=m+1$. 
    The Wigner function, a quasi-probability distribution in the oscillator phase space, is directly measured via photon number parity measurements after variable cavity displacements~\cite{vlastakis_deterministically_2013}. 
    From each Wigner function we reconstruct the density matrix, and the most significant off-diagonal element is $\rho_{n,m}$ (or $\rho_{n'm'}$) that reflects the coherence between $\ket{n}$ and $\ket{m}$ (or between $\ket{n'}$ and $\ket{m'}$).  We also perform similar measurements with permutations of odd-parity superpositions as input states (not shown).  The $\chi$ matrix block describing the coherence of the process can be computed by combining all the off-diagonal elements in these reconstructed density matrices (Extended Data Fig.~\ref{fig:process_matrix}).  The result for the six key elements characterizing PReSPA coherence, $\chi_{nm,n'm'}$, are shown next to the vertical arrows and to a good approximation equal to  $\rho_{n'm'}/\rho_{n,m}$.
    The deviations of $\chi_{nm,n'm'}$ from unity reflects the infidelity of the PReSPA process.}
\end{figure*}

Each of the two frequency combs is generated by single-sideband modulation of a microwave carrier source with four intermediate-frequency (IF) control signals digitally combined.  The amplitudes and phases of these IF signals can all be independently tuned.  The ancilla state is measured using dispersive readout via mode $R$, which is not sufficiently optimized to enable single-shot state assignment.  Measurement outcomes are converted to a nominal excited state probability by scaling the averaged demodulated signal relative to the reference ground and excited states. After a series of calibration experiments and compensating for multi-tone parametric mixing effects (see Methods), we experimentally obtain $\lambda\approx 27$ kHz, $\Omega\approx 88$ kHz for all transition paths.  

\begin{figure*}[tbp]
    \centering
    \includegraphics[scale=0.43]{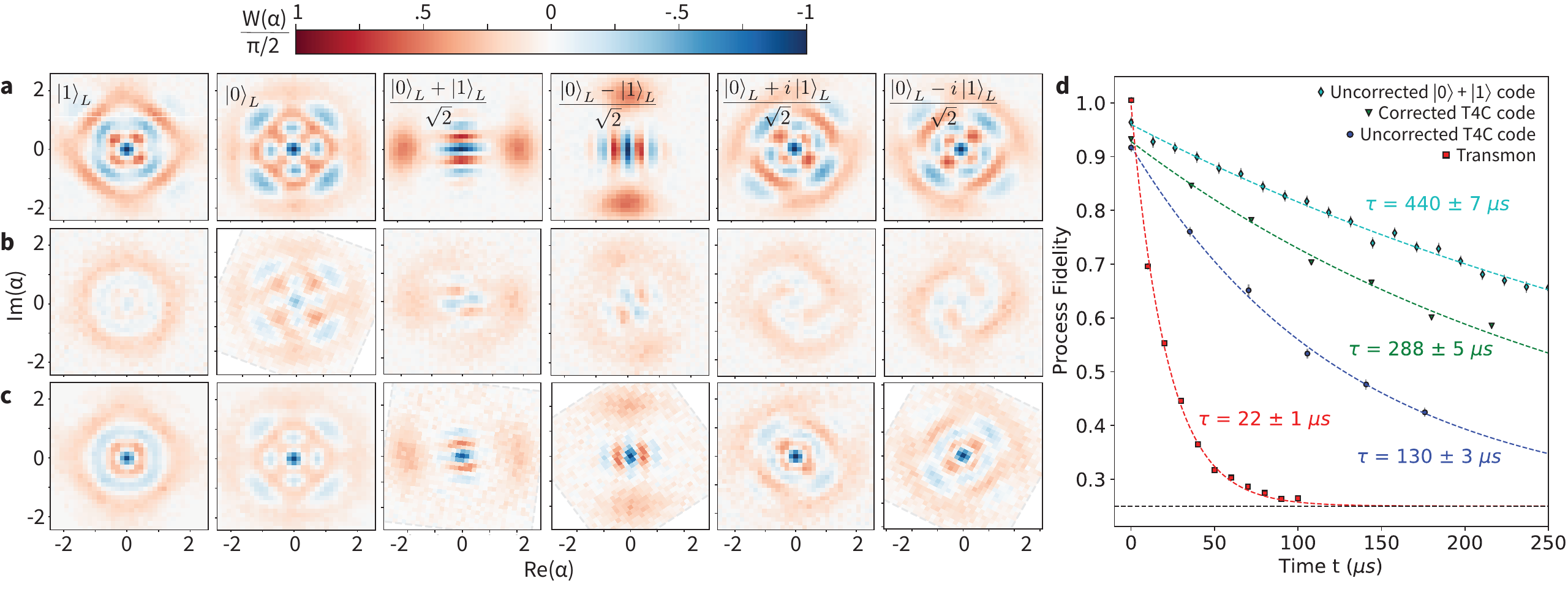}
    \caption{\textbf{AQEC performance.}  \textbf{a-c,} Cavity Wigner tomography for the six cardinal-point states of our Truncated 4-component Cat (T4C) code.  We measure the states at (\textbf{a}) $t=0$, (\textbf{b}) after 143 $\mu$s of free evolution, and (\textbf{c}) after 143 $\mu$s of AQEC using PReSPA. Center points of the Wigner function show a measurement of state parity, which is well-preserved under PReSPA.  The Wigner functions have been manually rotated into the correct rotating frame for the ease of comparison.  \textbf{d}, Quantum process fidelity of information storage with the AQEC protocol (green) compared with other reference methods performed with the same physical system.  
    We use each method to store quantum states corresponding to the six cardinal points of the Bloch sphere, wait for a variable time $t$ (or apply PReSPA for the AQEC curve), retrieve the state via the transmon using a decoding unitary, and perform quantum state tomography of the transmon.  Process fidelity is calculated from the quantum state fidelity for the corresponding six measurements.  All curves are fitted to the model $F(t) = .25 + A e^{-t/\tau}$ to extract characteristic times and error bars are one standard error.
    }
\end{figure*}

\subsection*{Characterization of dissipation operator}

We characterize the engineered PReSPA operator first by tracking the probability distribution of photon numbers in cavity $A$ over time.  This distribution can be measured using
spectroscopy of the ancilla qubit (Fig.~2a), whose frequency shifts by $-\chi_q$ for every additional photon~\cite{schuster_resolving_2007}.  As an example, Fig.~2b,c shows dissipative generation of a one-photon Fock state from vacuum under PReSPA as a result of the $\ket{1}\bra{0}$ element of the operator.  
To demonstrate the simultaneous action of the four parallel conversion paths, we apply PReSPA to an even-parity cat state ($\alpha=1.6$).
As expected, the cavity state converges to an odd-parity state, with all four Fock components shifted up by one photon (Fig.~2d,e).  The characteristic time scale of each conversion process can be measured by preparing the Fock state $\ket{2n}$ and tracking the population of $\ket{2n+1}$ as a function of time (Fig.~2f). 
The four conversion paths show well-matched temporal profiles with convergence half-time of $T_{eo}\approx 8$ $\mu$s.  The fidelity of the conversion decreases for higher photon numbers, 
consistent with faster photon loss.

To fully benchmark an engineered jump operator like PReSPA, a different type of tomographic characterization and experimental fidelity measure 
has to be devised. Here for simplicity, we probe the process $\chi$ matrix over $25$ $\mu$s ($\gg T_{eo}$) of PReSPA pumping as a proxy to experimentally benchmark the operator.  
The population conversion from $\ket{n}_A$ to $\ket{n'}_A$ is expressed in the $\chi_{nn,n'n'}$ matrix elements, which are measured using spectroscopy and shown in the Fig.~2f inset.

Quantum coherence of the even-to-odd conversion is reflected in the six key elements of the process matrix, $\chi_{02,13}$, $\chi_{24,35}$, $\chi_{46,57}$, $\chi_{04,15}$, $\chi_{26,37}$ and $\chi_{06,17}$, which can be measured from the transformation of off-diagonal elements of the relevant density matrices. 
We prepare the six pair-wise superposition states of $\ket{0}$, $\ket{2}$, $\ket{4}$ and $\ket{6}$ as input states, and measure the odd-parity output states at $t=25$ $\mu$s using Wigner tomography~\cite{vlastakis_deterministically_2013} (Fig.~3).  The interference fringes in the output Wigner functions demonstrate the persistence of coherence after photon addition.  From density matrix reconstruction, we find that PReSPA converts even-parity superposition states with nearly zero additional phase and 
an average fidelity of approximately 67\%.  


For the process $\chi$ matrix to reflect the full extent of the parity recovery, we choose a relatively long evolution time of 25 $\mu$s so that the initial even-to-odd conversion process is expected to complete with $>$98\% probability.  Various imperfections in the system also accumulate over time, including cavity dephasing, thermal excitations, and ongoing single-photon losses that trigger new rounds of PReSPA operation.  Therefore, the above characterization should be interpreted as measuring the combined process of a one-time 
PReSPA operation with a finite success probability together with continuous degradation of quantum states over 25 $\mu$s.  


\subsection*{AQEC performance}
We demonstrate AQEC by applying PReSPA to a set of six cavity multiphoton states corresponding to the cardinal points of the logical Bloch sphere. Over ~143 $\mu$s we measure the Wigner function of the cavity states demonstrating that  PReSPA preserves the Wigner negativity, a measure of non-classicality, to a greater extent than after uncorrected free evolution.

We can further retrieve the logical state from the cavity after a variable time with a unitary operation mapping it to the ancilla transmon, $\ket{\psi_L}\rightarrow x\ket{g}_q+ y\ket{e}_q$. 
We maximize this information retrieval by mapping both the logical code word, $\ket{0_L}$, as well as its orthogonal vector, $C_5\ket{1}-C_1\ket{5}$, to the same transmon state $\ket{g}_q$ (likewise for $\ket{1_L}$ and $\ket{e}_q$) to create a decoding that is insensitive to the size of the encoding. We observe a longitudinal relaxation time of $\Gamma_l^{-1}= 365\pm8$ $\mu$s for the pole states ($\ket{0_L}$ and $\ket{1_L}$) and a transverse relaxation time of $\Gamma_t^{-1}= 258\pm6$ $\mu$s for the equator states.  
From the process fidelity of this information storage process (compared to the identity matrix), we determine the corrected logical qubit's lifetime is $288\pm5$ $\mu$s, more than twice its uncorrected counterpart (Fig.~4b).  However, as in any QEC scheme, redundant storage of information incurs an overhead in decoherence, which in our case causes the uncorrected T4C qubit to decay over three times faster than the uncorrectable bare-cavity qubit (represented by the lowest-energy Fock states).  Therefore our corrected qubit's lifetime has not yet reached the QEC break-even point as in a landmark demonstration earlier using digital feedback~\cite{ofek_extending_2016}.

The performance of the current AQEC demonstration is limited by a number of experimental factors, and has substantial room for improvement. 
Our ancilla transmon has a spurious thermal excitation rate of 
1.4 ms$^{-1}$, which increases up to 2.0 ms$^{-1}$ during AQEC due to a combination of heating effect and intrinsic off-resonance excitations (see Supplementary Materials). 
Spontaneous transmon excitation can activate a sequential two-photon-gain process under PReSPA, leading to an incoherent logical bit-flip error (Extended Data Fig.~\ref{fig:transmon_heating}).  This effect accounts for $\sim$80\% of the observed $\Gamma_l$ and $\sim$50\% of the $\Gamma_t$ of the logical qubit, which can be 
significantly reduced~\cite{serniak_direct_2019, peterer_coherence_2015} with state-of-the-art filtering and shielding of the thermal environment.  


The rest of the logical qubit decoherence can be mostly attributed to the finite success probabilities ($S_l$ and $S_t$) of PReSPA in correcting single-photon loss. We define $S_i=1-{G_i}^{-1}$  $(i=l,t)$, where $G_i$ is the QEC gain factor between the natural photon-loss rate ($\bar{n}/T_{1A}\approx$ 1/150 $\mu$s$^{-1}$) and the part of the corrected logical qubit's error rate ($\Gamma_l$ or $\Gamma_t$) that can still be attributed to photon loss.  
Various imperfections can each be analyzed in terms of a correction failure rate that reduces $S_l$ and $S_t$.  We estimate $S_l=89\%$ and $S_t=76\%$, with prominent failure modes including a second photon loss 
or ancilla decay during a PReSPA operation, virtually-activated pumping processes, and higher-order cavity non-linearity (see Extended Data Table III and Supplementary Materials).  The QEC success probability should be understood as a hypothetical measure for a standalone PReSPA operation.  
Actual fidelity of a quantum state that experiences an error-and-correct cycle over a realistic time frame includes additional contributions from new errors at total rates of $\Gamma_l$ and $\Gamma_t$ after the initial error is corrected.
For example, a logical equator state incurring a photon loss can be expected to recover after 25 $\mu$s with fidelity $F\approx S_t e^{-t\Gamma_t} =69\%$.  This is consistent with the characterized $\chi$ matrix for PReSPA process, and is further confirmed in a separate experiment using initialized error states (Supplementary Materials).  The required QEC success probability to reach break-even in the absence of any undesirable transmon excitation is $\frac{1}{3}S_l+\frac{2}{3}S_t > 1-\frac{2}{3}\bar{n}^{-1}=81\%$.  
Modest improvement in ancilla and cavity lifetimes can lift $S_l, S_t$ sufficiently above this threshold to enable AQEC above break-even under realistic experimental conditions (see Methods for simulation results).

\subsection*{Outlook}

The four-component cat code (and its truncated variants)~\cite{ofek_extending_2016, hu_quantum_2019, reinhold_error-corrected_2020, ma_error-transparent_2020} has been pursued as a hardware-efficient paradigm for universal quantum computation~\cite{mirrahimi_dynamically_2014}, offering a full gate set and first-order error protection for logical qubits encoded in single cavities. 
However, there has been a caveat: the repetitive parity checks required to correct single-photon loss~\cite{ofek_extending_2016, hu_quantum_2019} are not simultaneously compatible with the continuous driven dissipation needed for phase-space stabilization
~\cite{leghtas_confining_2015, mundhada_experimental_2019}.  A competing bosonic QEC approach based on two-component cat qubits with strongly-biased noise channels~\cite{lescanne_exponential_2020, grimm_stabilization_2020} defers the challenge of photon-loss correction to a next-level repetition code (\textit{e.g.~}with a chain of cavities), but comes at the cost of increased hardware complexity.  
Our introduction of PReSPA provides a non-invasive method for photon-loss correction, paving the road for concurrent logic operations or dissipative stabilization within the odd-parity subspace.  For example, modest four-photon dissipation~\cite{mirrahimi_dynamically_2014,mundhada_experimental_2019} induced by a second reservoir can be applied together with PReSPA to evacuated entropy from cat-size changes and correct for
phase drifts, hence completing fully-passive first-order protection of a single-cavity logical qubit.
Autonomously corrected quantum gates can be constructed from path-independent unitary operations acting on the code-space and error-space in parallel~\cite{ma_path-independent_2020, ma_error-transparent_2020, reinhold_error-corrected_2020}. 
To improve fault tolerance, forward propagation of ancilla relaxation errors can be suppressed  
by error-transparent processes through the second excited state of the transmon~\cite{reinhold_error-corrected_2020} or using a biased-noise ancilla~\cite{puri_stabilized_2019,grimm_stabilization_2020}. In addition, our cascaded pumping technique to realize path erasure 
can be generalized to construct a broad family of dissipation operators of the form $\hat{L}=\sum_{i,j}\lambda_{i,j}\ket{i}\bra{j}$ in the Fock basis, providing the tools for realizing various other AQEC schemes~\cite{lihm_implementation-independent_2018,albert_pair-cat_2019,kapit_hardware-efficient_2016}. 

For reducing errors in quantum computing, the development of intrinsically protected physical qubits by Hamiltonian engineering~\cite{doucot_physical_2012, gyenis_experimental_2019} and implementation of QEC codes based on redundancy~\cite{lidar_quantum_2013,knill_theory_1997} have often been considered two distinct pursuits. 
The establishment of continuous AQEC of prominent errors, joined by recent developments of driven qubits with biased noise channels~\cite{lescanne_exponential_2020,grimm_stabilization_2020}, bridges this divide. 
In our demonstration, the cavity qubit is protected by a QEC code executed by a driven-dissipative environment in quasi-equilibrium, as described by a time-independent rotating-frame Hamiltonian and dissipation operators.  
Beyond specific implementations in circuit QED, our work suggests reservoir engineering as a unifying force that applies the flexibility of code-based error correction to improve the robustness of physical qubits with minimal resource overhead.
\\

\textit{Acknowledgements} --  
We thank MIT Lincoln Lab for providing the Josephson traveling wave parametric amplifier (TWPA)~\cite{macklin_nearquantum-limited_2015} for our measurement. We thank Dario Rosenstock, Ebru Dogan, and Xiaowei Deng for assistance with the experiment. This research was supported by the U.S. Air Force Office of Scientific Research (FA9550-18-1-0092) and the Army Research Office (W911NF-17-1-0469 and W911NF-19-1-0016).\\

\textit{Author Contributions} -- 
J.M.G.~carried out the device design, microwave measurements, and data analysis of the experiment under the supervision of C.W.   B.B.~generated the numerical pulses for unitary control in the experiment under the supervision of J.K. J.L.~fabricated the device and contributed to the cryogenic preparation of the apparatus.  S.S.~carried out numerical simulations for the experiment.  B.B., J.K., and C.W.~developed the approximate AQEC theory. C.W.~conceived and oversaw this project.  J.M.G., B.B., J.K.~and C.W.~wrote the manuscript with input from all authors.\\

Correspondence and requests for materials should be addressed to Chen Wang, wangc@umass.edu.



\section*{Methods}
\renewcommand\thefigure{\arabic{figure}}  
\setcounter{figure}{0}
\renewcommand{\figurename}{\textbf{Extended Data Fig.}}
\renewcommand{\tablename}{\textbf{Extended Data Table}}

\subsection*{Approximate AQEC and decoding unitary}

\begin{figure*}[tbp]
    \centering
    \includegraphics[scale=.28]{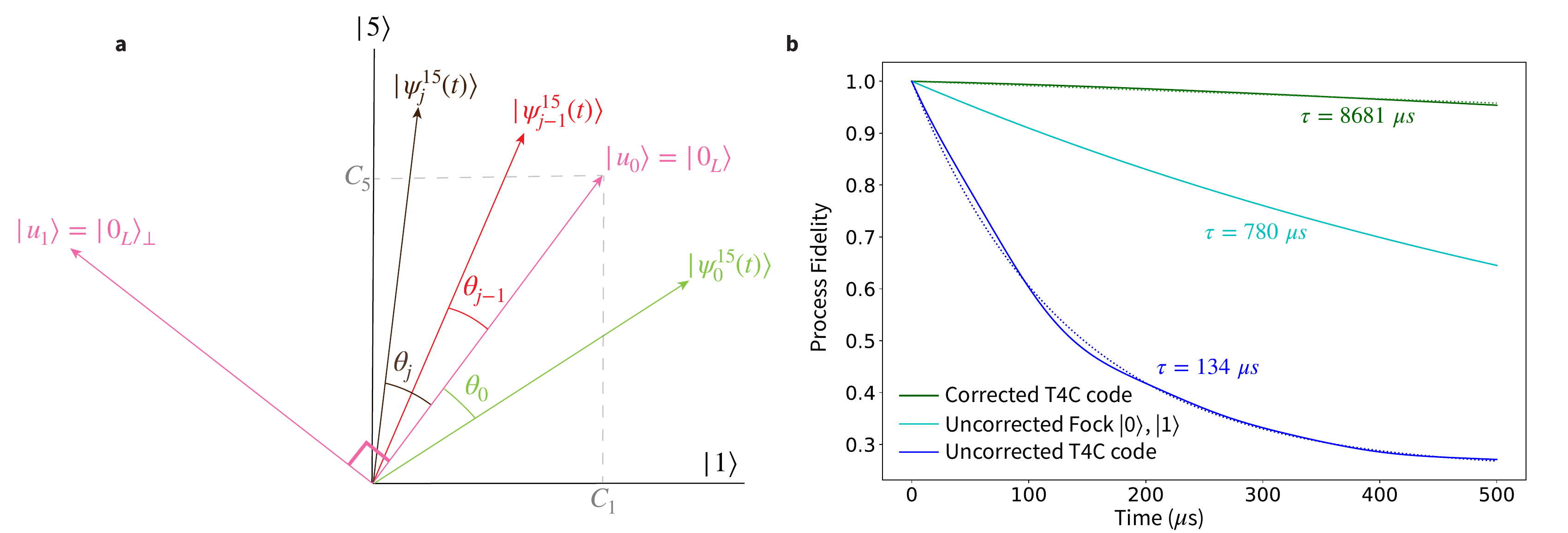}
    \caption{\textbf{a}, Diagram illustrating the quantum-trajectory state $\ket{\psi_j^{15}(t)}$ after a time period $t$, given a certain number $j$ of jumps. The initial state is $\ket{\psi(0)}=\ket{0_L}$ in the subspace $\mathcal{H}_{15}=\text{span}\,\{\ket{u_0},\ket{u_1}\}$. Both jump and no-jump evolution lead to rotation of $\ket{\psi_j^{15}(t)}$ over time within the subspace, and the angle $\theta_j(t)$ parametrizes this rotation.  For each quantum trajectory, $\ket{\psi_j^{15}(t)}$ slowly and continuously rotates clockwise in the absence of jumps and occasionally undergoes stochastic jumps clockwise.
    The diagram and the dynamics for the states $\ket{\psi_j^{37}(t)}$ in the $\mathcal{H}_{37}=\text{span}\,\{\ket{v_0},\ket{v_1}\}$ subspace (not shown) follows an analogous pattern. 
    \textbf{b}, Comparing the decay of process fidelities for three cases: T4C encoding using the ideal PReSPA scheme of this section (corrected T4C code, green), T4C encoding without using PReSPA (uncorrected T4C code, teal), and Fock state encoding (uncorrected Fock $\ket{0}$, $\ket{1}$, blue).  Experimental values of $T_{1A}$ and $K$ are used, and cavity dephasing is not considered.  Exponential curves for the T4C fidelity use the equation $\mathcal{F}_{\text{process}}(t) = 0.75e^{-t/\tau} + 0.25$ to extract decay rate $\tau$.  
    }
    \label{fig:decoding_vis_and_fidelity}
\end{figure*}

The T4C code belongs to the class of bosonic code with $N=2$ rotational symmetry~\cite{grimsmo_quantum_2020}.  
It can be understood as an approximate version of a Schr\"odinger cat code~\cite{leghtas_hardware-efficient_2013, mirrahimi_dynamically_2014} at small photon numbers.  In the idealized limit of large photon numbers and instantaneous operations of $\hat{\Pi}'_{eo}=\sum_{n=0}^\infty \ket{2n+1}\bra{2n}$ upon photon loss, 
such a cat code would enable perfect retrieval of quantum information.  Here we have extended the PReSPA operator to an infinite series $\hat{\Pi}'_{eo}$, which can be viewed as a parity-selective version of the Susskind-Glogower bare raising operator~\cite{susskind_quantum_1964}.  Practical limitations, however, necessitate a compromise accounting for the growth of the photon loss rate with average photon number and the increasing technical challenge of expanding the terms in the dissipation operator. The compromise struck with the T4C code employing the code words Eq.~(\ref{eq:code}) composed of odd-numbered Fock states $|n\rangle$ with $n\le7$.

Within T4C encoding, the joint effect of loss and instantaneous PReSPA, as described by a combined loss operator $\hat{\Pi}_{eo}\hat{a}$, generally induces an unwanted distortion of the original quantum state. Specifically, given a number of $j$ loss/PReSPA events within a time period $t$, the initial state $|\psi_0\rangle = x|0_L\rangle + y |1_L\rangle$ evolves into the state
\begin{equation}\label{final_state}
    \ket{\psi_j(t)} = x n^{15}_j(t)\ket{\psi_j^{15}(t)} + y n^{37}_j(t)\ket{\psi_j^{37}(t)}.      
\end{equation}
Here, $\ket{\psi_j^{kl}(t)} = \sum_{n=k,l}C_nn^{j/2}e^{-nt/2T_{1A}}\ket{n}/[n_j^{kl}N_j(t)]$  are orthonormal, 
and the coefficients recording the unwanted state distortion are
 $n^{kl}_j(t) = \| \sum_{n=k,l}C_n n^{j/2}e^{-nt/2T_{1A}}\ket{n}\| /N_j(t)$ where $k,l=1,5$ or $3,7$.

The decoding transformation seeks to transfer the amplitudes $x$, $y$ from the logical cavity qubit to the auxiliary transmon.
%
%
%
%
Perfect transfer is impeded by two key factors. First, the coefficients  $n_j^{kl}(t)$ in Eq.\ \eqref{final_state} will generally differ and, hence, lead to the mentioned distortion of $x$ and $y$. Second, the code words  undergo rotations within the $\mathcal{H}_{15}$ and $\mathcal{H}_{37}$ subspaces described by
\begin{align}\label{trajectory}
    \ket{\psi_j(t)} &= xn^{15}_j(t)\big(\cos\theta_j\ket{u_0}  
    + \sin\theta_j\ket{u_1}) \\\nonumber &+ yn^{37}_j(t)\big(\cos\varphi_j\ket{v_0} + \sin\varphi_j\ket{v_1}\big).
\end{align}
Here, the angles $\theta_j$ and $\varphi_j$ depend both on time and the number of loss events.  See Extended Data Fig.~\ref{fig:decoding_vis_and_fidelity}a for a visual description of the effect of jump and no-jump dynamics. The states $\{\ket{u_0}, \ket{u_1}\}$ and $\{\ket{v_0}, \ket{v_1}\}$ denote orthonormal bases of the $\mathcal{H}_{15}$ and $\mathcal{H}_{37}$ subspaces, respectively. 

Constructing a decoding unitary $\hat{U}_d$ such that:
\begin{align}
    \ket{g}\otimes\ket{u_0} \mapsto \ket{g0},\,\,\, \ket{g}\otimes\ket{u_1} \mapsto \ket{g1},\nonumber\\ \ket{g}\otimes\ket{v_0} \mapsto \ket{e0},\,\,\,\, \ket{g}\otimes\ket{v_1} \mapsto \ket{e1},
    \label{eq:decoding_U}
\end{align}
one obtains the reduced transmon density matrix
\begin{equation}
    \hat{\rho}_q =
    \sum_jp_j\begin{pmatrix}
    |x|^2(n^{15}_j)^2 & x^*y n^{15}_jn^{37}_j\cos \zeta_j \\ xy^*n^{15}_jn^{37}_j\cos \zeta_j & |y|^2(n^{37}_j)^2
    \end{pmatrix},
    \label{eq:decodedrho}
\end{equation}
with $p_j$ denoting the probability for observing $j$ loss events, and $\zeta_j = \theta_j-\varphi_j$.  In general, $\theta_j$ and $\varphi_j$ will not be identical, thus causing a second contribution to amplitude distortions. Nevertheless, they vary in the same direction as $j$ varies, and this partial cancellation plays a significant role in limiting the intrinsic loss of information in this approximate AQEC protocol.  See Supplementary Material for analytic details.

We further note that the photon-number expectation value, when averaged over all trajectories, remains exactly constant under the effect of $\hat{\Pi}_{eo}\hat{a}$.  In essence, despite its continuous decrease (under no-jump evolution) and stochastic increase (due to PReSPA after a photon loss), the effective cat size $|\alpha|$ of the encoding remains a constant on average.  This is different from the cat states in parity-measurement-based bosonic QEC experiments~\cite{ofek_extending_2016, reinhold_error-corrected_2020}.  In these experiments, the cat states decrease deterministically in size over a time scale of $1/2\kappa$, hence quickly experiencing wavefunction overlap between $C_{\alpha}$ and $C_{i\alpha}$.  The T4C code thus has the advantage of being self-sustained in energy, and the loss of orthogonality at small $\alpha$ is only encountered through a (2$^{nd}$-order) diffusive process.

We quantify deviations of Eq.~(\ref{eq:decodedrho}) from the intended target transmon states with the process fidelity $\mathcal{F}$ obtained by averaging $\mathcal{F}_{xy}(t) = |\langle \psi_q | \hat{\rho}_q(t) | \psi_q \rangle |^2$  over the six cardinal Bloch sphere points and rescaling its range to $1/4\le \mathcal{F} \le 1$. 
Extended Data Fig.~\ref{fig:decoding_vis_and_fidelity}b compares the theoretical process fidelity for PReSPA-corrected T4C (using the codeword coefficients in our experiment as shown below Eq.~(1)) to free evolution without PReSPA, for both T4C and Fock-state encoding. 
Theoretically, the corrected logical qubit incurs no longitudinal relaxation, and the transverse relaxation rate $\Gamma_t$ is 40 times lower than the single-photon loss rate $\bar{n}\kappa$, corresponding to an intrinsic QEC failure rate of $1-S_t=2.5\%$.

The T4C code can also be viewed as an instance of the binomial code class~\cite{michael_new_2016}, assuming the two logical codes contain equal photon numbers ($\bar{n}$).  Indeed, informational leakage associated with photon loss is minimized with equal $\bar{n}$ for $\ket{0_L}$ and $\ket{1_L}$, and the slight imbalance in our experimental code ($\bar{n}=3.6, 3.4$ for $\ket{0_L}$, $\ket{1_L}$) was incidental and not optimal.  However, we calculated that a revision to an equal $\bar{n}$ of 3.5 would only improve $S_t$ by 0.08\%, which is extremely marginal compared with all the imperfections that lead to $1-S_t\approx24\%$ in the experiment.  Assuming balanced $\bar{n}$ in a binomial code, the value of $\bar{n}$ can be chosen to balance the trade-off between the intrinsic QEC failure rate (minimized at $\bar{n}=4$) and photon-loss rate (minimized at $\bar{n}=3$, the smallest possible in this code).  Given the coherence parameters of the transmon and the cavity in this present study, the AQEC performance can be improved by choosing a smaller $\bar{n}$ close to 3.


\subsection*{Device and fabrication}
Our device uses a 3D-planar-hybrid cQED architecture~\cite{axline_architecture_2016}, and the design specifics are similar to Ref.~\onlinecite{wang_schrodinger_2016}.  The cavity is machined from 5N5 Al (99.9995\% pure) and houses a large waveguide section containing two cylindrical re-entrant quarter-wave resonators~\cite{reagor_quantum_2016} and a small waveguide tunnel to fit a sapphire chip. The transmon and the low-Q readout resonator are made from thin-film aluminum deposited on the sapphire chip. The transmon contains a single Al-AlO$_x$-Al Josephson junction and is fabricated using a Dolan bridge technique. Electron-beam lithography is carried out with a 30keV JEOL JSM-7001F SEM, and the evaporation/oxidation is performed with a Plassys MEB550S evaporator. 

\begin{table}[bp]
\caption{\textbf{System parameters.}\\ $^\dagger$The transmon $T_2^*$ reflects any $1/e$ decay time of Ramsey oscillations.  The transmon displays a random switching behavior between two values of $\omega_q/2\pi$ 40 kHz apart, with a dwell-time split of approximately 85\%:15\%.  The switching time scale is on the order of sub-seconds to seconds.  All our experimental data reflects the averaged result from sampling the two ancilla frequencies. \\
$^\ddagger$ The cross-Kerr $\chi_{Ar}$ is derived from other measured parameters. \\
$^\#$Cavity A has a distinctive switching behavior between a regular state with stable $T_{2A}$ ($380\pm25$ $\mu$s) and occasional ``bad periods" lasting for 2-8 hours where $T_{2A}$ fluctuates wildly in the range of 200-340 $\mu$s.  The reduced cavity coherence during these periods is not accompanied by any other changes of system parameters, and can be recovered by a Hahn echo pulse (with echo, cavity $T_{2A}\approx$ 390 $\mu$s at all times).  We exclude data during these unstable periods throughout the paper except for Extended Data Fig.~\ref{fig:fidelity_track} where its impact on AQEC performance is illustrated.}
\centering
\begin{tabular}{c c c c c}
\hline\hline\\[-2ex]
		& Symbol	&  Value	 \\
\hline\\[-2ex]
Transmon frequency  & $\omega_q/2\pi$	& 5489.59 MHz\\
Transmon anharmonicity  &  $\alpha_q/2\pi$	& 201.22 MHz \\
Transmon $T_1$  & $T_{1q}$	& $39$ $\mu$s \\
Transmon $T_2^*$ Ramsey  & $T^*_{2q}$ & $17$ $\mu$s$^\dagger$ \\
Transmon $T_2$ Echo  & 	& $36$ $\mu$s \\
Transmon $\ket{e}_q$ population  & & 5\% \\
\hline\\[-2ex]
Reservoir frequency  & $\omega_r/2\pi$ & 7337.8 MHz \\
Reservoir-transmon coupling  & $\chi_r/2\pi$ & 2.8 MHz \\
Reservoir $T_1$  & $1/\kappa$ & $0.27$ $\mu$s \\
\hline\\[-2ex]
Cavity $A$ frequency & $\omega_A/2\pi$ & 4067.445 MHz \\
Cavity $A$-transmon coupling & $\chi_q/2\pi$ & 1.313 MHz \\
Cavity $A$-reservoir coupling & $\chi_{Ar}/2\pi$ & 7 kHz$^{\ddagger}$\\
Cavity $A$ anharmonicity & $K/2\pi$ & 1.7 kHz \\
Cavity $A$ 2nd order coupling & $\chi_q '/2\pi$ & 5.5 kHz \\
Cavity $A$ $T_1$ & $T_{1A}$ & $520$ $\mu$s \\
Cavity $A$ $T_2$ & $T_{2A}$ & $380$ $\mu$s$^{\#}$ \\
Cavity $A$ $\ket{1}$ population &  & $\sim$1\% \\

\hline
\end{tabular}
\label{table:parameters}
\end{table}

Of the two high-Q storage cavity modes, the lower-frequency mode $A$ is used for this experiment while the higher-frequency (6.089 GHz) mode, $B$, is idle in the vacuum state throughout the experiment.  Cavity mode $B$ has a dispersive coupling with the transmon of 6.5 MHz that we found too high for implementing PReSPA. 
It has a thermal population of 1\%-2\%, which has a small contribution to the dephasing of $A$.

\subsection*{Driven Josephson circuit Hamiltonian under PReSPA}
The Hamiltonian of the circuit QED system can be derived using the perturbation theory~\cite{nigg_black-box_2012} that adds weak anharmonicity to three harmonic oscillator modes corresponding to the Cavity $A$, transmon ancilla $q$ and the stripline resonator $R$: 
\begin{align}
    \frac{\hat{H}}{\hbar} = \tilde{\omega}_q \hat{q}^\dagger \hat{q} + \tilde{\omega}_A \hat{a}^\dagger \hat{a} + \tilde{\omega}_R \hat{r}^\dagger \hat{r} - \frac{E_J}{\hbar} \bigg(\cos\hat{\varphi}+\frac{\hat{\varphi}^2}{2}\bigg)
    \label{eq:bareH}
\end{align}
where $\tilde{\omega}_q$, $\tilde{\omega}_A$, $\tilde{\omega}_R$ are the frequencies of the eigenmodes of the linearized system, $\hat{q}$, $\hat{a}$ and $\hat{r}$ their lowering operators,

\begin{figure*}[tbp]
    \centering
    \includegraphics[scale=.75]{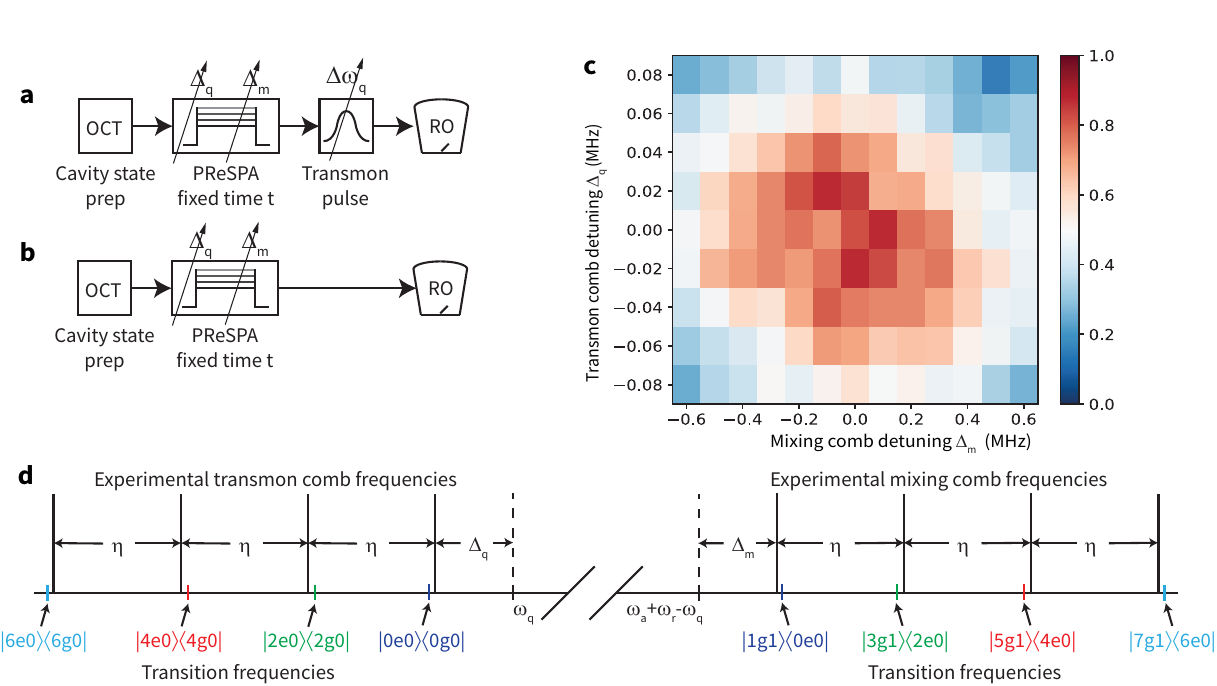}
    \caption{\textbf{PReSPA spectroscopy.} \textbf{a, b,} Control pulse sequence for two-dimensional (2d) spectroscopy to find the resonance conditions for the PReSPA mixing comb and transmon comb. We prepare an even-parity Fock state ($\ket{0}$, $\ket{2}$, $\ket{4}$, or $\ket{6}$), apply PReSPA for a fixed time (12 $\mu s$) with varying detunings of the transmon comb ($\Delta_q$) and the mixing comb ($\Delta_m$) in an attempt to activate dissipative photon addition.  After a 1 $\mu$s wait time for the reservoir to relax, we either (\textbf{a}) selectively $\pi$-pulse the transmon 
    conditioned on Cavity $A$ being in the targeted final state ($\ket{1}$, $\ket{3}$, $\ket{5}$, or $\ket{7}$)   
    or (\textbf{b}) skip this pulse (for a background measurement), and proceed to read out the transmon state. The difference between the two measurements informs the likelihood of successful photon addition.  
    \textbf{c}, 2d PReSPA spectroscopy data: likelihood of photon addition as a function of the comb detunings ($\Delta_q$ and $\Delta_m$) for the $\ket{0}$ to $\ket{1}$ transition.  Note that the linewidth of the four-wave-mixing transition is an order of magnitude greater than that of the transmon excitation due to the short reservoir $T_{1r}$.  We can repeat this procedure to find all four sets of transition frequencies. 
    \textbf{d}, Cartoon spectrum of PReSPA drive frequencies. Four transmon drives, left, and four mixing drives, right, compose PReSPA. The colored ticks indicate the actual transition frequencies while the vertical black bars show the microwave drive frequencies in PReSPA. The transmon drive for the $\ket{0}$ to $\ket{1}$ conversion process is approximately at the Stark-shifted transmon frequency, $\omega_q-
    \Delta_\text{Stark}$, and the $\ket{0}$ to $\ket{1}$ mixing drive is near $\omega_A + \omega_R - \omega_q+\Delta_\text{Stark}$. 
    Because of the equal frequency spacing $\eta$ in each comb and the unequal frequency spacing between the transitions with different photon numbers (due to the 6$^{th}$-order non-linearity, $\chi'_q$), not all drives can be placed exactly on resonance.  Experimentally, we settle for $\eta$ slightly greater than $2\chi_q$, and $\Delta_q=\Delta_m$ slightly smaller than $\Delta_\text{Stark}$ to compensate for the effect of $\chi'_q$. 
    \label{fig:spectroscpy}
    }
\end{figure*}

$E_J$ the Josephson energy of the junction, and $\hat{\varphi} = \phi_q \hat{q} + \phi_A \hat{a} + \phi_R \hat{r} + \mathrm{h.c.}$ the phase operator of the Josephson junction. $\phi_q$, $\phi_A$, $\phi_R$ are the mode zero-point phase fluctuations across the junction for modes $q$, $A$ and $R$.

Implementing PReSPA requires two frequency combs: one comb exciting the transmon, and one mixing comb converting that transmon excitation into excitations in $A$ and $R$. This mixing comb activates a four-wave mixing process using the nonlinearity of the Josephson junction. To account for the four mixing tones with frequencies $\omega_m$ in our Hamiltonian, we make a unitary transformation by displacing the qubit annihilation operator $\hat{q} \rightarrow \hat{q} + \sum_m\xi_m e^{i \omega_m t}$, and the phase across the junction is (see \textit{e.g.~}Supplementary materials of Ref.~\onlinecite{leghtas_confining_2015}): 
\begin{align}
    \hat{\varphi} = \phi_q \hat{q} + \phi_A \hat{a} 
    + \phi_R \hat{r} + 
    \sum_{\substack{m=0}}^3 \phi_q \xi_m e^{i \omega_m t} + \mathrm{h.c.}
    \label{eq:JJphase}
\end{align}
Expanding the cosine, going to a rotating frame, and employing a dispersive transformation that cancels 2nd order terms, the relevant 4th and 6th order terms include non-driven terms (without $\xi_m$) and driven terms. The non-driven terms, generic to most circuit QED systems, are:
\begin{align}\label{rotframe_ham}
    \frac{\hat{H}_\text{nd}}{\hbar} = &- \chi_q \hat{q}^\dagger \hat{q}\hat{a}^\dagger\hat{a}- - \chi_r\hat{q}^\dagger \hat{q}\hat{r}^\dagger \hat{r}  
    - \chi_{Ar}\hat{a}^\dagger\hat{a} \hat{r}^\dagger \hat{r}
     \nonumber \\  &- \frac{K}{2}\hat{a}^\dagger\hat{a}^\dagger\hat{a}\hat{a} 
    - \frac{\chi'_q}{2}\hat{q}^\dagger \hat{q}\hat{a}^\dagger\hat{a}^\dagger\hat{a}\hat{a}
\end{align}
where $\chi_q$, $\chi_r$, $\chi_{Ar}$ are the dispersive couplings between the transmon and Cavity $A$, between transmon and reservoir $R$, and between $A$ and $R$, respectively.  $K$ is the self Kerr of Cavity $A$, and $\chi'_q$ is the 6$^{th}$ order non-linearity.  We treat the transmon as a two-level system since the higher excited states are not accessed in this experiment.
Likewise, for all operations except readout, $R$ is either in the ground or first excited state and we ignore its higher order terms. 

By setting $\omega_m = \omega_a + \omega_r - \omega_q + m\eta$ we get stationary, or slowly rotating, four-wave mixing terms:
\begin{equation}
    \frac{\hat{H}_\text{mix}}{\hbar} = -\frac{E_J}{\hbar} \phi_q^2 \phi_a \phi_r \sum_{m=0}^3 \xi_m e^{im\eta t} \hat{q}\hat{a}^\dagger\hat{r}^\dagger + \mathrm{h.c.}
    \label{eq:Hmix}
\end{equation}
where $\eta$ is the difference in frequency between each nearest pair of mixing tones.
The four mixing tones will each individually Stark shift the transmon but we can simply absorb those into the transmon frequency for the rotating frame transformation mentioned above. There will also be slowly-rotating Stark shift terms that are a result of the cross terms of two different mixing tones:
\begin{equation}
    \frac{\hat{H}_\text{Stark}}{\hbar} = -\frac{E_J}{\hbar} \phi_q^4 \sum_{\substack{k=0}}^3 \sum_{\substack{l=0 \\ l\neq k}}^3  \xi_k \xi_l^* e^{-i(l-k)\eta t} \hat{q}^{\dagger}\hat{q}
    \label{eq:Hstark}
\end{equation}

The four mixing tones acting together drive the four $\ket{2n,e,0}\leftrightarrow\ket{2n+1,g,1}$ ($n=0,1,2,3$) transitions (the black solid arrows in Fig.~1c).  We can calculate the complex Rabi rate of these driven transitions under a weak-drive approximation of $|\xi_k|^2\ll \hbar\eta/(E_J\phi_q^4)$: 
\begin{align}
    &\Omega_{n} = -\frac{E_J}{\hbar} \sqrt{2n+1} \phi_q^2 \phi_a \phi_r \bigg[\xi_n-\frac{E_J \phi_q^4}{\hbar \eta}\sum_{m;k;l \neq k}\frac{\xi_m}{(l-k)}\nonumber\\
    &\times\left( \xi_k \xi_l^* \delta_{n-m,l-k} - \xi_k^* \xi_l\delta_{n-m,k-l} \right)  
    + \mathcal{O}( \frac{E_J^2\phi_q^8 \xi^4}{\hbar^2 \eta^2})\bigg] 
    \label{eq:OmegaCalc}
\end{align}
where $\delta$ is the Kronecker delta function.   This rate is caused by two different mechanisms: a direct drive by one of the four tones that is on resonance, and a multi-tone parametric effect. 
This parametric effect arises from the time-modulation of transmon frequency by the Stark shift induced by pairs of mixing tones. When the detuning of one of the off-resonant mixing tones exactly matches the modulation frequency, the transition can be parametrically driven by a combination of three tones. Depending on the relative phases of the three tones, these terms can contribute constructively or destructively to the transition rate.  Due to the bosonic enhancement factor $\sqrt{2n+1}$, higher photon number states typically require smaller $\xi_n$.

\begin{table*}[tbp]
\caption{\textbf{Comparison of calculated and measured PReSPA transition rates}.  Using calibrated amplitudes and phases of the transmon comb and mixing comb in the experiment, the complex-valued PReSPA transition rates ($\lambda_n$ and $\Omega_{n}$) can be calculated based on Eqs.~(\ref{eq:OmegaCalc}, \ref{eq:lambdaCalc}). Here the transmon comb amplitude $\Lambda_n$ is calibrated from ancilla Rabi oscillations, and the amplitudes of mixing tones are approximately converted to the dimensionless displacement parameter $\xi_n$ by measuring the Stark shift $\Delta_\text{Stark}$ induced by that single tone: $\xi\approx\sqrt{\Delta_\text{Stark}/2 \alpha_q}$.  The experimentally measured PReSPA transition rates result from fitting the time-domain dynamics of the photon addition processes as in Extended Data Fig.~\ref{fig:rates}, which do not contain phases.  Note that $\Omega_0$ is intentionally set opposite to others in phase to suppress multi-tone mixing effects [Eqs.~(\ref{eq:OmegaCalc}, \ref{eq:lambdaCalc})] by destructive interference.
}
\centering
\begin{tabular}{c c c c c c}
\hline\hline\\[-2ex]
index n	& raw mixing amp.~(AU)\,	& \, $\xi_n$\ (approx.)	&  \, $\Omega_{n}$ (kHz) (theory)\, & $\abs{\Omega_{n}}$ (kHz) (fit)	 \\
\hline\\[-2.8ex]
\\[-1em]
0 ($\ket{0}$ to $\ket{1}$) & -1.22  & -0.058        & -125   & 92\\
\\[-1em]
1 ($\ket{2}$ to $\ket{3}$) & \,1.00   & \,0.048     & \,127   & 88\\
\\[-1em]
2 ($\ket{4}$ to $\ket{5}$) & \,0.64    & \,0.030     & \,127   & 87\\
\\[-1em]
3 ($\ket{6}$ to $\ket{7}$) & \,0.49    & \,0.023     & \,124   & 85\\

\hline\hline\\[-2ex]
index n	& raw transmon amp.~(AU)\,	& $\Lambda_n$ (kHz) &  $\lambda_n$ (kHz) (theory)	&  $|\lambda_n|$ (kHz) (fit) 	& \\
\hline\\[-2.8ex]
\\[-1em]
0 ($\ket{0}$ to $\ket{1}$) & \,\,-0.98 $e^{-0.43i}$   & \, -21.4 $e^{-0.43i}$ & \,\,-27 $e^{-0.37i}$   &  28\\
\\[-1em]
1 ($\ket{2}$ to $\ket{3}$) & 1.52 $e^{0.00i}$   &  33.1 $e^{0.00i}$ & 28 $e^{0.04i}$    & 27\\
\\[-1em]
2 ($\ket{4}$ to $\ket{5}$) & 1.27 $e^{0.02i}$  &  26.7 $e^{0.02i}$\,   & 28 $e^{0.02i}$   &27\\
\\[-1em]
3 ($\ket{6}$ to $\ket{7}$) & \,\,\,\,1.14 $e^{-0.35i}$   & \,\, 24.9 $e^{-0.35i}$ & \,\,\,\,27  $e^{-0.34i}$   &26 \\
\hline
\end{tabular}
\label{table:PReSPA_rates}
\end{table*}

This calculation has been done with the assumption that the spacing between tones $\eta$ is constant but we could also consider non-even frequency spacing. In this case, terms that rotate very slowly with time would appear in the mixing rates which would cause our PReSPA operator to undesirably change with time. 
We make the choice of keeping $\eta$ constant to avoid these complications. 

A similar calculation can be done for the four selective transmon drives. To preserve path independence we use the same frequency spacing magnitude $\abs{\eta}$ as the mixing drives but with the opposite sign. As the transmon drives are relatively weak, they do not cause a significant single-tone or multi-tone Stark shift but the frequency modulation caused by the mixing drives will allow for similar three-tone parametric mixing effects (two mixing tones and one transmon tone). We can write the rates for the four transmon transitions as:
\begin{equation}
    \lambda_{n} = \Lambda_n -\frac{2 E_J \phi_q^4}{\hbar\eta} \sum_{m \neq n; k; l \neq k} \frac{\Lambda_m \xi_k \xi_l}{m-n} \delta_{n-m,k-l}
    \label{eq:lambdaCalc}
\end{equation}
where $\Lambda_n$ are the bare transmon Rabi rates without the mixing drives.  For either comb, the effect of this parametric mixing is a renormalization of the Rabi rates $\Omega_n$ and $\lambda_n$ from the Rabi rates by individual resonant tones (that are proportional to $\xi_n$ and $\Lambda_n$).\\

\subsection*{PReSPA spectroscopy and rates}  
Implementing PReSPA requires initial parameter guesses supplemented with experimental corrections. We make the initial assumption that the comb frequency spacing $\eta = 2\chi_q$ and that the $\ket{0e0}\leftrightarrow\ket{1g1}$ (mixing) transition frequency is $\omega_{A} + \omega_{R} - \omega_{q}$. Because our strong off-resonant drives Stark shift the transmon, a detuning $\Delta$ needs to be added to the mixing comb frequencies and subtracted from the transmon comb frequencies to best match the desired transitions.

\begin{figure*}[tbp]
    \centering
    \includegraphics[scale=.72]{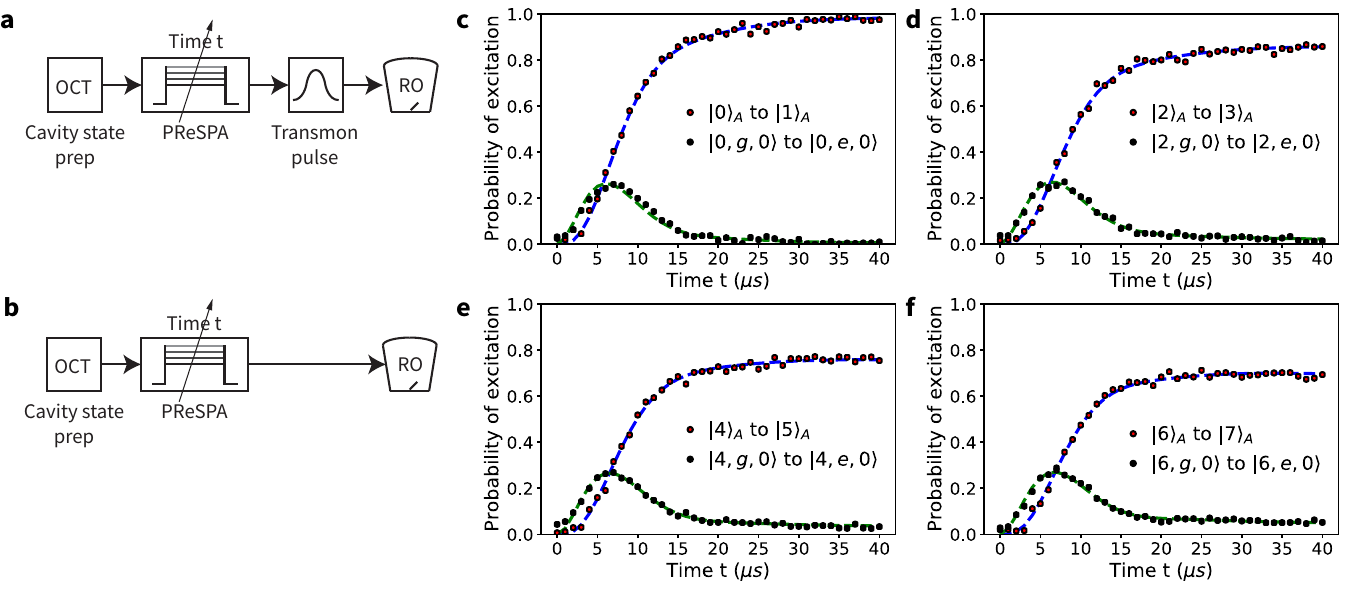}
    \caption{\textbf{Calibration of PReSPA transition rates. a}, \textbf{b}, Control pulse sequence diagram for tracking the convergence of cavity state towards a target Fock state and the transmon excitation probability over time.  
    For tracking the cavity state (\textbf{a}), we prepare initial even Fock states $\ket{0}$, $\ket{2}$, $\ket{4}$, and $\ket{6}$, perform a variable time of PReSPA, and a photon-number selective transmon $\pi$-pulse at fixed detunings ($-1\chi_q$, $-3\chi_q$, $-5\chi_q$, and $-7\chi_q$ respectively), and read out the transmon state.  For tracking transmon excitations (\textbf{b}), the $\pi$-pulse is omitted.  
    \textbf{c-f}, Transmon excitation probability (black dots) and cavity photon addition probability (red dots) calculated from these two measurement sequences.  These measurements are performed for photon conversion from each even state to the next corresponding odd state.  The dynamics of excitation transfer is not in the Zeno limit of $\kappa\gg\Omega,\lambda$, but rather can be understood as a slightly-overdamped coupled-oscillator system.  To include various detunings and decoherence rates, we fit these curves
    with a numerically-solved two-stage pumping model (blue and green dashed curves) and extract the PReSPA transition rates ($\Omega_n$ and $\lambda_n$).  The fit contains only three free parameters: $\Omega_n$, $\lambda_n$ and a y-axis scaling factor.  
    \label{fig:rates}
    }
\end{figure*}

\begin{figure*}[tbp]
    \centering
    \includegraphics[scale=.50]{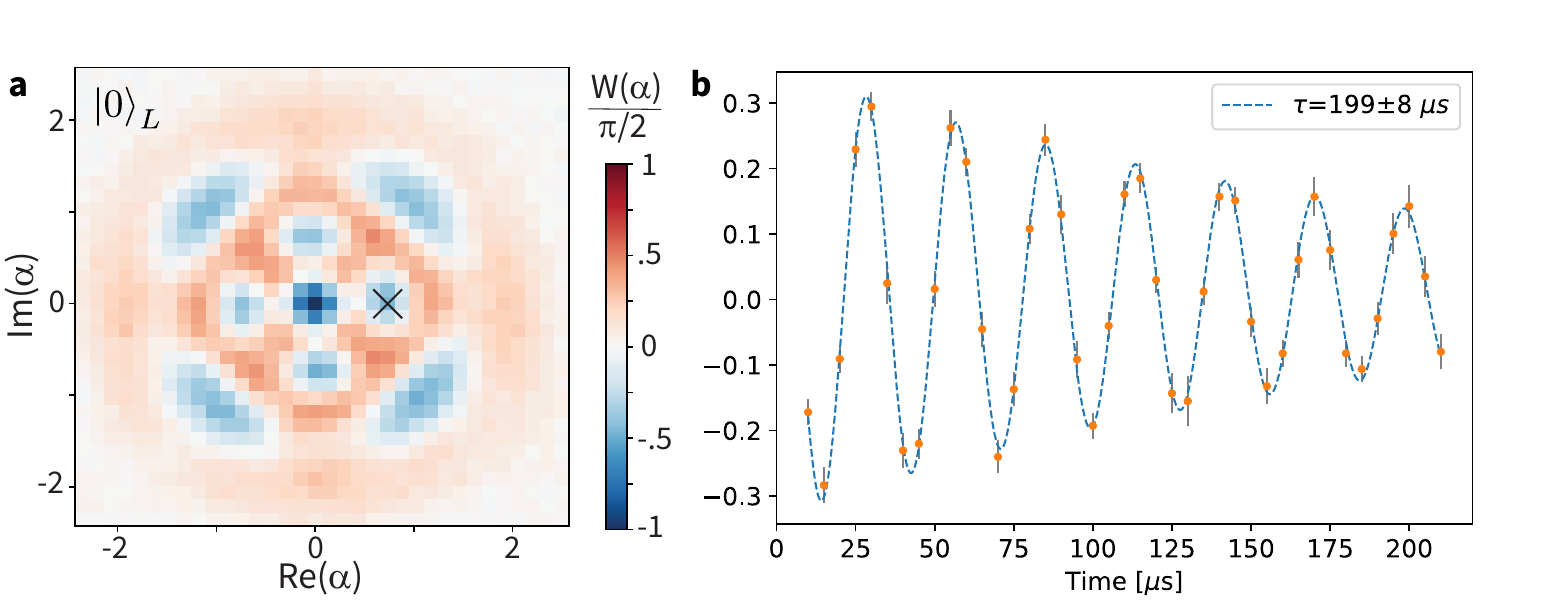}
    \caption{\textbf{Cavity Wigner and PReSPA Ramsey measurements. a}, Experimental Wigner function $W(\alpha)$ of $\ket{0_L}$, acquired by applying a cavity displacement operation $\hat{D}_{\alpha}=\exp(\alpha\hat{a}^\dagger-\alpha^*\hat{a})$ with variable complex amplitude $\alpha$ followed by an ancilla-assisted photon-number-parity measurement (which is composed of two $\pi/2$ pulses of the ancilla and a delay time of $\pi/\chi_q$ and an ancilla readout~\cite{vlastakis_deterministically_2013,sun_tracking_2014}).  The Wigner function rotates around the origin over time at a rate proportional to the frequency difference between $\ket{1}$ and $\ket{5}$ in the rotating frame of the experiment. \textbf{b}, Measured Wigner function values at a fixed phase-space position (as indicated by the cross in \textbf{a}, at $\alpha = 0.75$) as a function of time under PReSPA.  Analogous to a qubit Ramsey measurement, this cavity PReSPA Ramsey experiment can be used to efficiently track the phase evolution of any two-component superposition states using the interference effect enabled by the coherent cavity displacement ($\hat{D}_{\alpha}$) before readout.  The exponential envelope of the sinusoidal fit indicates the rate of decay for the coherence between $\ket{1}$ and $\ket{5}$ under the correction of PReSPA.  
    Similar measurements are applied to various superposition states to provide direct calibration of the frequencies and phases of these states under PReSPA. PReSPA enhances the ability to use such Ramsey measurements at high photon numbers since it approximately preserves photon number distributions in the cavity.
    \label{fig:ramsey}
    }
\end{figure*}

In addition, the presence of non-zero $\chi'_q$ adds a quadratic shift to the transmon frequency based on the cavity photon occupation.  Therefore, there is no choice of evenly-spaced tones that can drive all transitions on resonance.  To best match the experimentally measured transitions we calibrate $\eta$ and $\Delta$ through a set of spectroscopy measurements (Extended Data Fig.~\ref{fig:spectroscpy}c).  Experimentally we choose $\eta = 2.679$ MHz and $\Delta \approx 2.9$ MHz (calibrated on a daily basis) to ensure no pair of tones are farther than 10 kHz ($\approx 2\chi'_q$) off-resonant to their corresponding transitions. 

To match the dissipative processes across the four parallel paths in PReSPA, we measure the probability of photon addition, over time, to each of the four even states as described in Extended Data Fig.~\ref{fig:rates}. By fitting the curves of photon population and transmon excitation probability we can extract $\Omega_n$ and $\lambda_n$ for each of the four conversion paths. The fit model considers the quantum dynamics across the four relevant levels ($\ket{2n,g,0}$, $\ket{2n,e,0}$, $\ket{2n+1,g,1}$ and $\ket{2n+1,g,0}$) in a two-stage pumping process. 
It is simulated with QuTiP and includes the small detuning of the drives due to $\chi'_q$, the decay of Cavity $A$ and Reservoir $R$, and the relaxation, dephasing, and the spurious two-frequency switching behavior (see notes in Extended Data Table I) of the transmon. 

From Eqs.~(\ref{eq:OmegaCalc}, \ref{eq:lambdaCalc}) we can also estimate $\Omega_n$ and $\lambda_n$ from the amplitudes and phases of the microwave tones.  We find experimentally that achieving equal $\Omega_n$ and $\lambda_n$ for the four conversion paths requires significantly different microwave amplitudes within each comb, in quantitative agreement with theoretical predictions (Extended Data Table II).  There is a discrepancy in a global pre-factor in the magnitude of $\Omega_n$, possibly caused by the coarse estimates of zero point fluctuation parameters ($\phi's$) from measurable experimental parameters (\textit{i.e.} frequencies and dispersive shifts).

\subsection*{Cavity and transmon Ramsey under PReSPA}
To characterize the phase evolution of various multiphoton superposition states under PReSPA, we perform a Ramsey-style experiment (PReSPA Ramsey, as shown in Extended Data Fig.~\ref{fig:ramsey}). This can be used both for measuring the frequency  and coherence of odd-parity superposition states under PReSPA and for tuning up superposition phases in the PReSPA operator $\hat{\Pi}_{eo}$.

We measure the phase imparted in the photon addition process by starting with an equal superposition of two even states (\textit{e.g.~}$\frac{\ket{0}+\ket{2}}{\sqrt{2}}$) and perform a PReSPA Ramsey at times greater than 25 $\mu$s, sufficient for nearly-complete even-to-odd conversion.  By fitting the Ramsey oscillations, we can extract the phase of the PReSPA-converted state (\textit{e.g.~}$\frac{\ket{1}+\ket{3}}{\sqrt{2}}$), which is calibrated against the phase extracted from another PReSPA Ramsey experiment on the corresponding odd-parity initial state (\textit{e.g.~}$\frac{\ket{1}+\ket{3}}{\sqrt{2}}$).  
Phase and amplitude parameters of the microwave combs are adjusted to minimize the imparted phases of the photon addition process. 

PReSPA Ramsey is further used to measure the phase accumulation rate of our logical code words. When measuring the process fidelity of error correction we work in a frame where $\ket{0}_L$ is stationary. By measuring the phase of $\ket{1}_L$ as a function of time we can calculate the correct decoding angle.

\begin{figure*}[tbp]
    \centering
    \includegraphics[scale=.35]{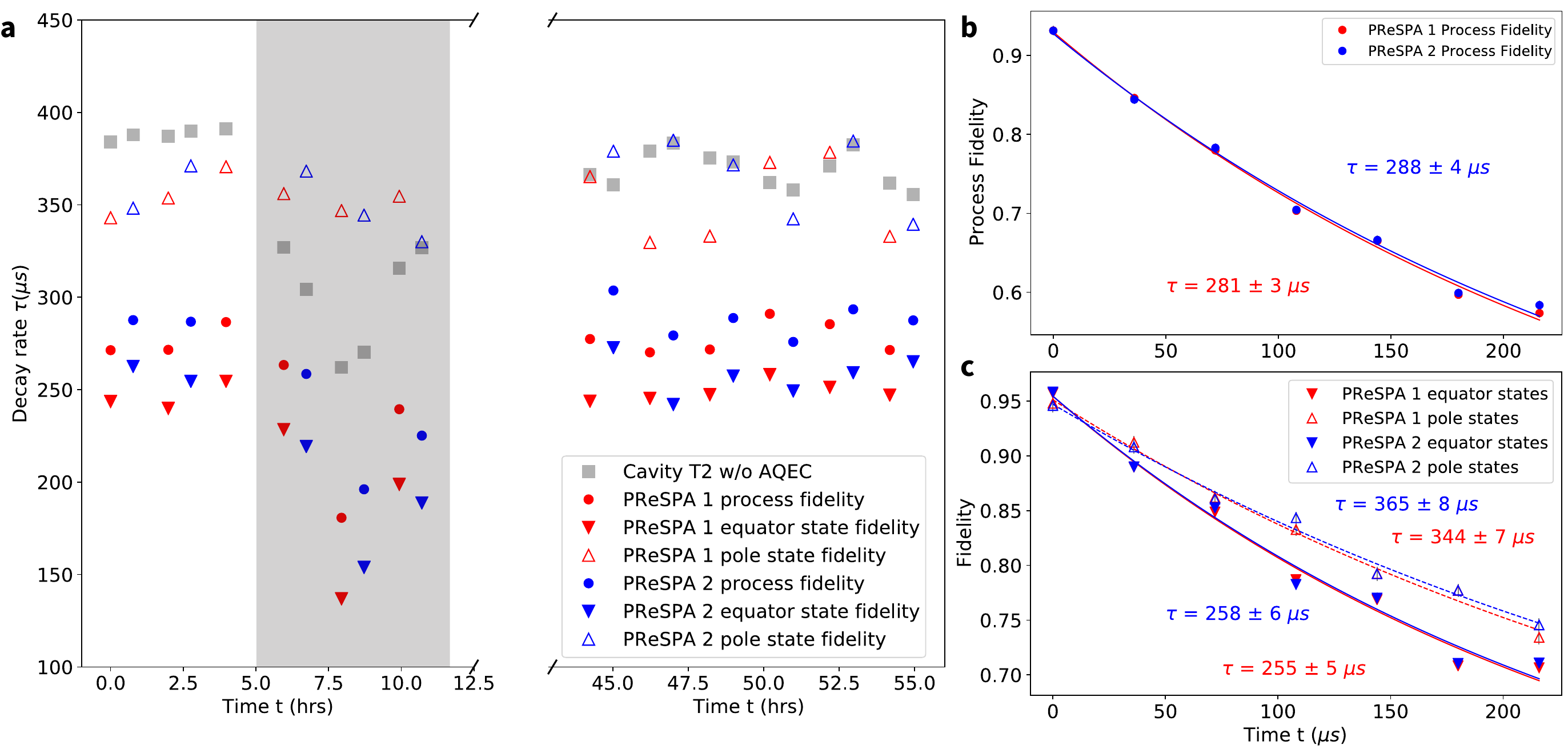}
    \caption{\textbf{Tracking AQEC performance over time.} \textbf{a}, We interleave measurements of the corrected logical qubits under two differently-calibrated PReSPA parameter sets (PReSPA-1, red and PReSPA-2, blue) while also monitoring the cavity $T_{2A}$. Each circle corresponds to the decay time of process fidelity extracted from measuring all six cardinal points of the logical Bloch sphere as described in Fig.~4.  The decay rates for state fidelity are shown in triangles for the two logical pole states (upwards triangles) and the four equator states (downwards triangles). 
    The state fidelity is measured by quantum state tomography of the ancilla after decoding the cavity state as described in  Methods.  For ancilla state tomography, we measure all six Pauli operators ($\hat{\sigma}_x$, $\hat{\sigma}_y$, $\hat{\sigma}_z$, $-\hat{\sigma}_x$, $-\hat\sigma_y$, $-\hat\sigma_z$) by performing ancilla rotations before readout. The over-complete measurement set is used for simultaneous calibration of the readout signal contrast, allowing for accurate determination of the transmon state.
    PReSPA-1 is calibrated by adjusting control parameters to achieve matched PReSPA rates and zero conversion phases as discussed in the Methods. For PReSPA-2, we employ further empirical parameter optimization to maximize equator state lifetime as described in the Supplementary Materials.  Cavity $A$ has a two-state switching behavior of unknown origin (see notes in Extended Data Table I).  For distinct stretches of 2-8 hours, Cavity $A$ shows fluctuating and abnormally low $T_{2A}$, and all data recorded during such periods (with example data shown in the shaded region) are excluded in all other parts of the paper. 
    \textbf{b}, Process fidelity averaged over the data, excluding the shaded region, for both PReSPA 1 and 2. Data reported in Fig.~4d for the corrected T4C encoding is a duplication of the blue points here. \textbf{c}, Equator and pole state fidelity for the same time period for both PReSPA 1 and 2. Oscillatory behavior in the data is caused by the numerical differences between the two decoding pulses discussed in ``GRAPE Methods."
    \label{fig:fidelity_track}
    }
\end{figure*}

We also perform a regular qubit Ramsey experiment for the transmon in the presence of the mixing combs of PReSPA (without exciting Cavity $A$).  This allows measurement of the Stark shift of the transmon rapidly and accurately, because to a very good approximation, only the mixing comb contributes to the Stark shift of the transmon. 
This experiment further shows that the coherence time of the transmon is not affected by the strong off-resonant drives of PReSPA (with $T^*_{2q}=17$ $\mu$s).

\subsection*{GRAPE methods}  
Pulses for state preparation and decoding are constructed using quantum optimal control (QOC) theory \cite{werschnik_quantum_2007,glaser_training_2015,gollub_monotonic_2008,heeres_implementing_2017}, as adapted for arbitrary cavity-state controls in circuit QED in Ref.~\onlinecite{heeres_implementing_2017} (see Supplementary Materials, Sec.\ 2.4 for additional details). QOC  numerically optimizes the envelopes of tones acting on the transmon and storage cavity, such that deviations of the realized state or unitary from the target state or target unitary are minimized. Deviations are quantified as state or process infidelities and incorporated into a cost functional $C$, which is subject to gradient-descent minimization via GRAPE (Gradient Ascent Pulse Engineering) \cite{khaneja_optimal_2005}.

Two envelope-modulated drives with carrier frequencies $\omega_A$ and $\omega_q$ are applied to storage cavity and transmon for preparation of the cavity states shown in Figs.\ 3a and 4a. By adjusting the respective envelopes, GRAPE maximizes the state-transfer fidelity $\mathcal{F}_{\text{prep}}= |\bra{\psi_T}\ket{\psi}|^2$, 
where $\ket{\psi_T}$ and $\ket{\psi}$ are the cavity target state and the actual cavity state realized by the pulse. Closed-system time evolution is simulated based on the rotating-frame Hamiltonian \eqref{rotframe_ham}. 

GRAPE is further utilized in constructing a decoding unitary $\hat{U}_d$ seeking to map the T4C-encoded cavity state to the transmon.  As minimal constraints on $\hat{U}_d$, it is required to act on the four relevant basis states as shown in Eq.~(\ref{eq:decoding_U}).  This does not fully determine the unitary in the seven-dimensional subspace $\mathcal{S}$ spanned by $\ket{g,n}$ and $\ket{e,\ell}$ where $n = 0,1,3,5,7$ and $\ell = 0,1$. We therefore specified three additional constraints for the numerical optimization. 
However aside from preserving unitarity, these additional constraints were discretional. 
Beyond unitarity, no constraints were placed in the algorithm on the action of $\hat{U}_d$ on states in the remaining space orthogonal to $\mathcal{S}$.  In this study, the actual form of transformation on those states is not characterized and may have a small positive or negative impact on the fidelity of decoded qubit state.  Future enforcement of additional constraints may lead to small improvements in logical fidelity.

\begin{table*}[tbp]
\caption{\textbf{Breakdown of the decoherence sources of the logical qubit under AQEC.} See Supplementary Materials for detailed discussions on all error channels.  
Some of the listed decoherence channels can be unambiguously associated with discrete quantum jumps, such as transmon $\gamma_\uparrow$, $T_{1q}$ error, 2$^{nd}$ photon loss and other cavity dephasing.  Each event fully scrambles cavity photons phases and therefore translate to logical transverse relaxation at a 1:1 ratio.   Bit-flip rate from the occurrence of double photon gain ($\sim$50\% of $\gamma_\uparrow$ events, see Extended Data Fig.~\ref{fig:transmon_heating}) and 2$^{nd}$ photon loss is converted to longitudinal relaxation rate via a factor 2 by definition. 
Other decoherence channels are more continuous in nature, for which we define their effective rate of occurrence as the resultant logical transverse relaxation rate.  They arise from various competing time scales in PReSPA such as dispersive shift against reservoir linewidth and dissipation rates against various higher-order nonlinearity.  The scaling law for each contribution is also listed when applicable. $\gamma_m$ is the transmon decay rate via the reservoir in the presence of the mixing comb, $\sim$0.35 ms$^{-1}$ in our experiment.  Definition of other rates can be found in Extended Data Table I or in the main text. 
For intrinsic code word distortion, see Methods-Approximate AQEC and decoding unitary.  
Other cavity dephasing effects are caused by thermal excitation in the reservoir mode and other cavity modes of the system. 
Other possible small PReSPA phase errors include imperfect matching of $\lambda$ and $\Omega$ for the four photon addition paths, and second-order sensitivity to a low-frequency-noise problem present in our transmon ancilla (see Extended Data Table I caption).}
\centering
\begin{tabular}{c c c c c}
\hline\hline\\[-2ex]
	& Rate of occurrence\,	& \,  Longitudinal relaxation rate\,	&  \, Transverse relaxation rate	&  Rate scaling  \\
\hline\\[-2ex]
\textbf{Unrelated to single-photon loss:} \\
ancilla spurious excitation $\gamma_\uparrow$ &   1.8 ms$^{-1}$ & 1.9 ms$^{-1}$  & 1.8 ms$^{-1}$ &  \\
off-resonant excitation by transmon comb &   0.2 ms$^{-1}$ & 0.3 ms$^{-1}$    & 0.2 ms$^{-1}$   & $<\gamma_m(\lambda/\chi_q)^2$\\
other cavity dephasing 	&  0.3 ms$^{-1}$	& & 0.3 ms$^{-1}$ & \\
\hline\\[-2ex]
\textbf{Fail to correct single-photon loss:} & \\
ancilla relaxation $T_{1q}$         &  $7\%\cdot(\bar{n}/T_{1A})$ &   & 0.5 ms$^{-1}$  & $1/(\gamma_{m}T_{1q})\cdot(\bar{n}/T_{1A})$\\
$2^{nd}$ photon decay	            & $6\%\cdot(\bar{n}/T_{1A})$ & 0.7 ms$^{-1}$ & 0.4 ms$^{-1}$ & $(\bar{n}t_{eo}/T_{1A})\cdot(\bar{n}/T_{1A})$ \\
incorrect pumping path by mixing comb
& $3\%\cdot(\bar{n}/T_{1A})$ & & 0.2 ms$^{-1}$ & $(\kappa/2\chi_q)^2\cdot(\bar{n}/T_{1A})$ \\
$K$ \& correction time uncertainty	& $2\%\cdot(\bar{n}/T_{1A})$  & & 0.1 ms$^{-1}$ &  $(K t_{eo})^2\cdot(\bar{n}/T_{1A})$ \\
$\chi'_q$ \& correction time uncertainty    & $1\%\cdot(\bar{n}/T_{1A})$ & & 0.1 ms$^{-1}$ & $(\chi'_q/\gamma_{m})^2\cdot(\bar{n}/T_{1A})$\\
$\chi_{Ar}$ \& correction time uncertainty  & $<0.5\%\cdot(\bar{n}/T_{1A})$& &  & $(\chi_{Ar}/\kappa)^2\cdot(\bar{n}/T_{1A})$\\
intrinsic code word distortion	    & $2.5\%\cdot(\bar{n}/T_{1A})$ & & 0.2 ms$^{-1}$ & \\
other PReSPA phase errors	        & $\sim2\%\cdot(\bar{n}/T_{1A})$ & & $\sim0.1$ ms$^{-1}$ & \\
\hline\\[-2ex]
Total	& & 2.8 ms$^{-1}$	&  3.8 ms$^{-1}$  &   \\[1ex]
\hline
\end{tabular}
\label{table:errors}
\end{table*}

\subsection*{Cavity Kerr effect on decoding}

The truncated cat states evolve with a deterministic collapse-revival cycle due to the self Kerr introduced by the Josephson junction~\cite{kirchmair_observation_2013}.  This effect does not directly cause logical qubit decay (except for a small reduction to QEC success probability $S_t$, see Extended Data Table III).  However, it imposes an inconvenience that for an arbitrary time $t$ (modulo $\pi/K$ or 288 $\mu s$), a dedicated OCT decoding pulse is needed to account for the phase accumulation due to the Kerr.  In our experiment, by accounting for an effective $Z$-rotation of the logical qubit after decoding, we can recycle the same decoding pulse every $\pi/4K$ or 72 $\mu s$. 

To see this, in the rotating frame chosen to make $\ket{0_L}$ stationary (calibrated by the PReSPA Ramsey, Extended Data Fig.~\ref{fig:ramsey}), the Kerr evolution of a T4C state (without decoherence) is: 
\begin{align}
    \ket{\psi} = x\big(C_1\ket{1} 
    + C_5\ket{5}\big)
    + y e^{-2iKt} \big( C_3 \ket{3} 
    + C_7 e^{8iKt} \ket{7}\big)
\end{align}
In the experiment, we computed two different decoding pulses.  The first decoding pulse, used for $t=0$ (mod 72 $\mu$s), implements the original transformation prescribed by Eq.~(\ref{eq:decoding_U}):
\begin{align}
    \ket{g}\otimes\big(C_1\ket{1} 
    + C_5\ket{5}\big) \mapsto \ket{g0},\nonumber\\ \ket{g}\otimes\big(C_1\ket{5} 
    - C_5\ket{1}\big) \mapsto \ket{g1},\nonumber\\ \ket{g}\otimes( C_3 \ket{3} 
    + C_7 \ket{7}\big) \mapsto \ket{e0},\nonumber\\ \ket{g}\otimes( C_3 \ket{7} 
    - C_7 \ket{3}\big) \mapsto \ket{e1},
    \label{eq:decoding_U2}
\end{align}
The second decoding pulse, used for $t=36$ $\mu$s (mod 72 $\mu$s), implements a modified transformation where the coefficient of $\ket{7}$ acquires a extra minus sign in Eq.~(\ref{eq:decoding_U2}).  
    
These two decoding pulses differ slightly in their performance.  This is responsible for the alternating pattern in the state and process fidelity of T4C code (as a function of time) in Fig.~4b and Extended Data Fig.~\ref{fig:fidelity_track}b, c.

\subsection*{Simulation of experimental AQEC}
In order to verify the error contribution from various experimental imperfections, we perform numerical simulations of the AQEC experiment by solving the master equation in QuTiP.  The simulation is carried out in a Hilbert space of $9\times2\times2$ dimensions and a rotating frame of frequencies $\omega_q$, $\omega_A$, $\omega_r$ for the transmon, cavity and reservoir excitations. The fast dynamics on the scale of mode frequencies are therefore not simulated, but the relative dynamics of different photon-number states on the time scale of $\sim\chi_q$ are fully captured.

\begin{figure*}[tbp]
    \centering
    \includegraphics[scale=.5]{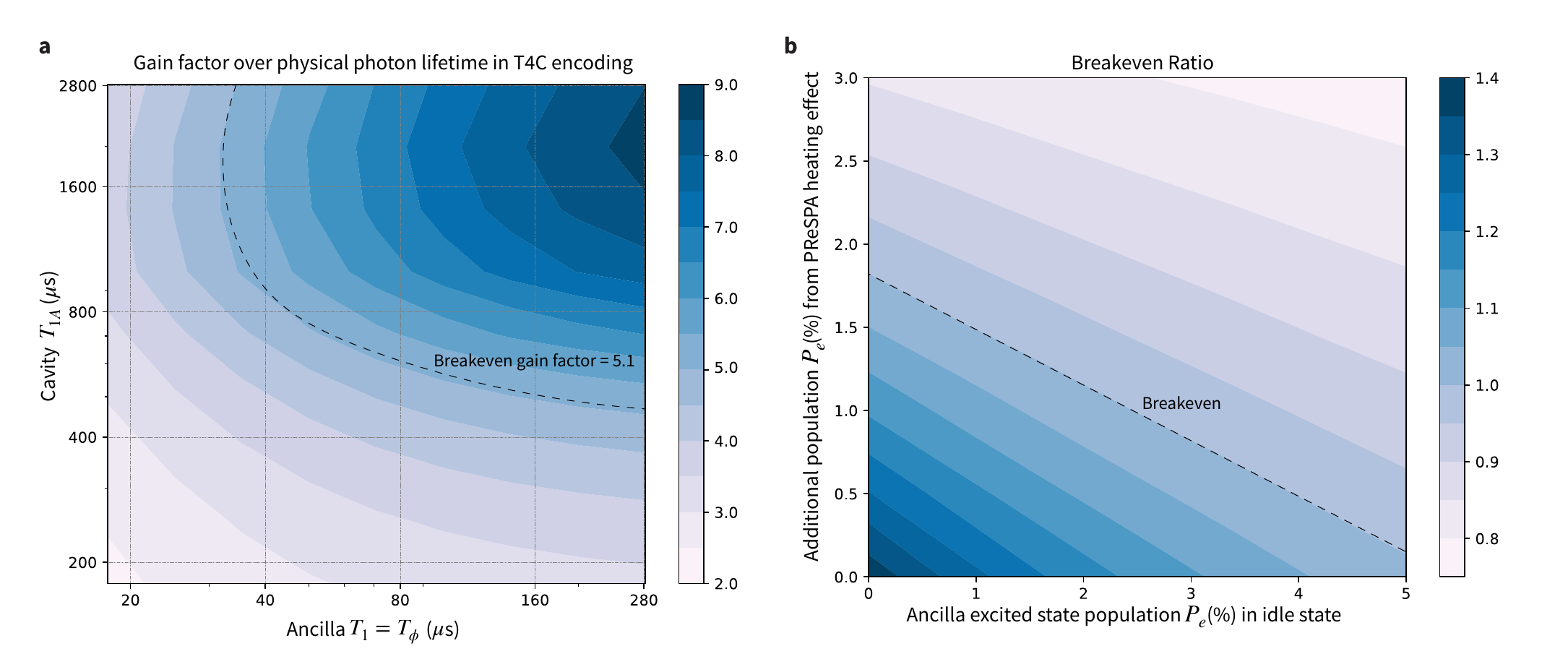}
    \caption{\textbf{Predicted AQEC performance in numerical simulations.}  The results are based on master-equation simulations of a T4C qubit (with encoding $\bar{n}=3.4$ for both basis states) under the Hamiltonian Eqs.~(\ref{rotframe_ham}) and (\ref{Drive_hamilt}) which captures the dynamics under the microwave combs of PReSPA.
    \textbf{a,} Gain factor of the corrected logical qubit lifetime over the physical photon lifetime ($T_{1A}/\bar{n}$) in the T4C encoding as a function of Ancilla $T_{1q}$, $T_{\phi}$ (which are made equal for convenience) and cavity $T_{1A}$.  To illustrate the intrinsic performance of our transmon-based PReSPA pumping scheme, we have assumed no ancilla thermal excitations and other cavity dephasing errors.  However, ancilla excitations due to the imperfect frequency selectivity of PReSPA, which is unrelated to photon loss, is reflected in the simulation.  Therefore, the gain factor shown is different from $G_i$ defined in the main text, and decreases slightly at long $T_{1A}$. 
    \textbf{b,} The QEC break-even ratio, defined as the T4C qubit lifetime under PReSPA over the lifetime of the $\ket{0/1}$ Fock-state encoding.  Here we use a specific set of achievable coherence times $T_{1A}=1$ ms,  $T_{1q}=T_{\phi}=100$ $\mu$s~\cite{wang_schrodinger_2016, heeres_implementing_2017} and show the degradation of AQEC performance in the presence of spontaneous transmon excitation ($\gamma_\uparrow$) errors caused by the stray thermal background (horizontal axis) or pump-tone-induced heating from PReSPA (vertical axis).  QEC breakeven can be reached if the $\gamma_\uparrow$ rate is kept reasonably low.  The dashed lines in both \textbf{a} and \textbf{b} indicate where the QEC break-even ratio equals 1.\\
    Relevant system parameters: In \textbf{a}, we use $\lambda/2\pi=17.5$ kHz, $\Omega/2\pi=45$ kHz, $\kappa/2\pi=227$ kHz, $\chi_q/2\pi=1.05$ MHz, $\chi_{r}/2\pi=1.6$ MHz, scaled from the experiment by 50\%, 50\%, 40\%, 80\%, 80\% respectively.  In \textbf{b}, we use $\lambda/2\pi=21.6$ kHz, $\Omega/2\pi=72$ kHz, $\kappa/2\pi=364$ kHz, $\chi_q/2\pi=1.18$ MHz, $\chi_{r}/2\pi=1.8$ MHz, scaled from the experiment by 80\%, 80\%, 64\%, 90\%, 90\% respectively.  The choice of parameters here is guided by the scaling laws of various error channels (Extended Data Table III) but did not go through optimization of individual parameters.  
    \label{fig:gain}
    }
\end{figure*}

The simulated system Hamiltonian is composed of the non-driven terms $\hat{H}_{nd}$ (Eq.~(\ref{rotframe_ham}) with parameters listed in Extended Data Table~\ref{table:parameters}) and the following time-dependent drive Hamiltonian: 
\begin{align}\label{Drive_hamilt}
     \hat{H}_{d}/\hbar= &\sum_{n=0}^{3}e^{-i[2n\chi_{q}+(5n-2)\chi_{q}']t}\sum_{m=0}^{8}(\lambda_n\ket{m,e,0}\bra{m,g,0} \nonumber \\ &+ \Omega_n\sqrt{\frac{m+1}{2n+1}}\ket{m,e,0}\bra{m+1,g,1}) + \text{h.c.}
\end{align} 
This drive Hamiltonian describes four transmon excitation tones and four four-wave mixing tones that are $\pm2\chi_{q}'$ detuned from the transition frequencies, as shown in Extended Data Fig.~\ref{fig:spectroscpy}. 
$\lambda_n$ and $\Omega_n$ are the transmon drive rate and four-wave mixing rate for the intended transitions as listed in Extended Data Table II, accounting for the small differences in rates within each comb.  Crucially, Eq.~(\ref{Drive_hamilt}) encapsulates the resonant and off-resonant dynamics of each photon-number state ($m$ = 0 to 8) under the effect of all four tones ($n$ = 0 to 3) in each PReSPA comb.  We assume the constant Stark shift of all modes have been included in the rotating frame.  We also make the assumption that the effect of two-tone modulated Stark shift is fully captured by the renormalization of comb rates $\lambda_n$ and $\Omega_n$ in the spirit of Eq.~(\ref{eq:OmegaCalc}) and (\ref{eq:lambdaCalc}).

In addition to single photon loss, we account for the spurious excitations, relaxation, dephasing of the ancilla and other causes of cavity dephasing. The QuTip simulation thus solves the master equation, 
\begin{align}
     &\pdv{\rho(t)}{t}= -\frac{i}{\hbar}\Big[(\hat{H}_{\text{nd}}+\hat{H}_{d}), \rho(t)\Big] + \Big(\frac{1}{T_{1A}}D[\hat{a}]+\frac{1}{T_{1r}}D[\hat{r}]\nonumber\\
     &+\frac{1}{T_{1q}}D[\hat{q}] + \frac{1}{T_{\phi}}D[\hat{q}^{\dagger}\hat{q}]+ \gamma_{\uparrow}D[\hat{q}^\dagger] + \gamma_{\phi}D[\hat{a}^{\dagger}\hat{a}] \Big)\rho(t)
     \label{eq:master_eqn}
\end{align} 
where $D[\hat{O}]\rho(t) = \hat{O}\rho\hat{O}^{\dagger} -\frac{1}{2} \{\hat{O}^{\dagger}\hat{O}, \hat{\rho}\}$. The $T_{1A}$, $T_{1r}$, $T_{1q}$ are the relaxation times for cavity, reservoir and transmon respectively. $\gamma_{\uparrow}$ is the spurious ancilla excitation rate and $\gamma_{\phi}$ is the storage cavity dephasing rate.

The density matrix of the system at any given time $t$ is transformed by the decoding unitary Eq.~\eqref{eq:decoding_U} so that the cavity logical information is decoded into the transmon state just like in the experiment.  The cavity Kerr effect is taken into account by adjusting the phases of the decoding unitary as discussed earlier in Methods.  The quantum state fidelity against the targeted ancilla state is then computed as a function of time $t$. 

The simulation shows a corrected logical qubit lifetime of 290 $\mu$s, with $\Gamma^{-1}_l=383$ $\mu$s and $\Gamma^{-1}_t=255$ $\mu$s, in excellent agreement with the experiment.  We also performed sensitivity analysis to individual Hamiltonian or loss parameters in the simulation, and the results are consistent with our estimates of individual error channels presented in Extended Data Table III and the Supplementary Materials.

\begin{figure*}[tbp]
    \centering
    \includegraphics[scale=.63]{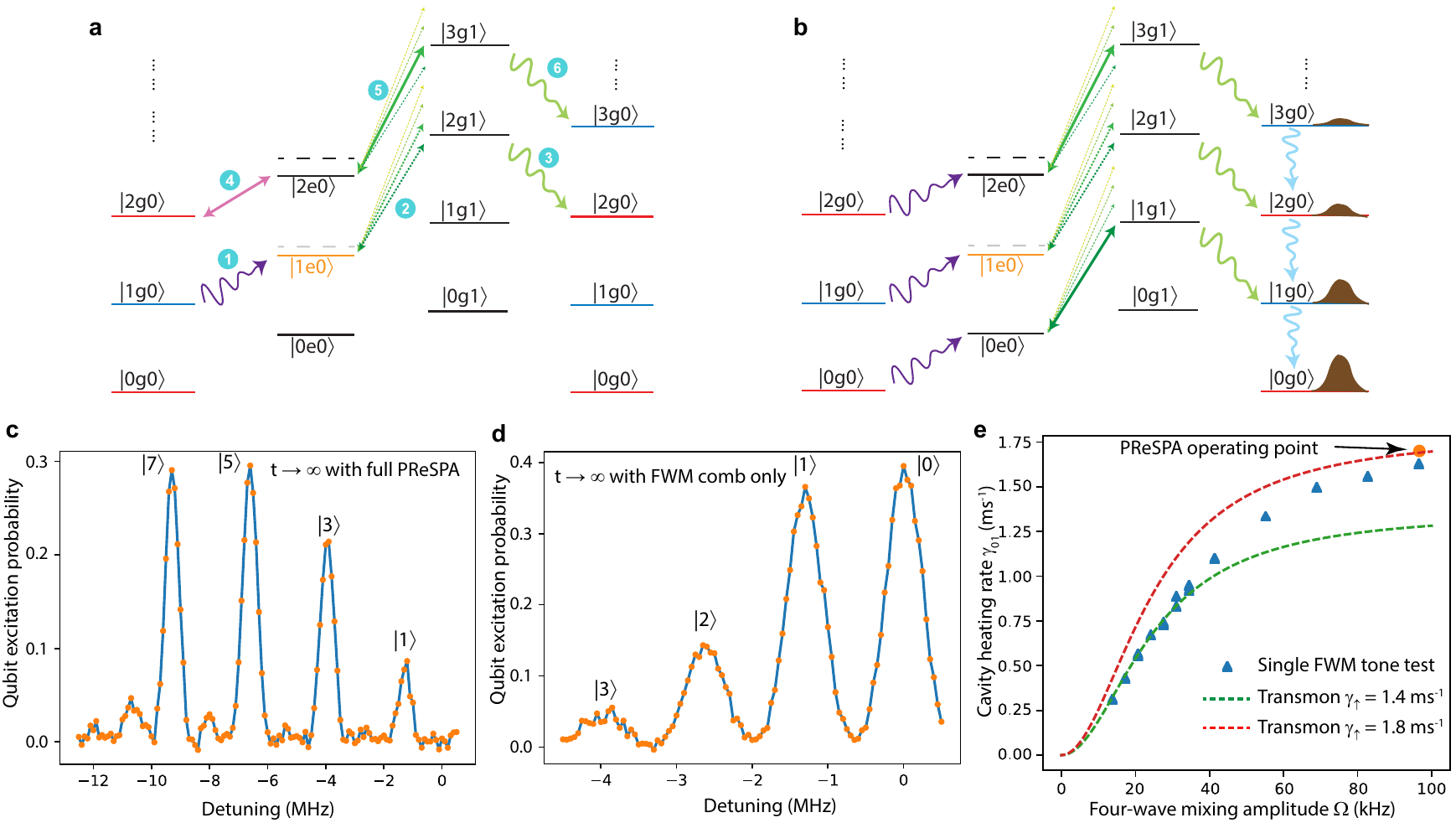}
    \caption{\textbf{Cavity heating effect caused by spurious transmon excitations.} 
    \textbf{a}, Schematic mechanism of sequential two-photon gain triggered by spurious excitation of the transmon.  Starting from $\ket{1g0}$, following the transitions labeled with circled numbers, the system is excited to $\ket{1e0}$ by a transmon $\gamma_\uparrow$ jump, and then driven unintentionally to $\ket{2g1}$ by mixing tones that are off-resonance by only $\pm\chi_q/2\pi=\pm1.3$ MHz (which does not provide strong enough frequency selectivity relative to the reservoir linewidth $\kappa/2\pi=0.58$ MHz), and then relaxes to $\ket{2g0}$ following reservoir decay.  Once a photon is added in this spurious odd-to-even conversion process, the PReSPA scheme by design will add a second photon, driving the system ultimately to $\ket{3g0}$.  Similarly, a transmon $\gamma_\uparrow$ jump can add two photons to $\ket{3}$ and $\ket{5}$ states (but not $\ket{7}$).
    \textbf{b}, Schematic illustration of the steady-state photon number distribution established between the ancilla-$\gamma_\uparrow$-induced photon addition and the natural photon loss of the cavity.  The figure corresponds to the configuration of the test experiment in (\textbf{d}) when only the mixing comb (but not the transmon excitation comb) is applied.   
    \textbf{c}, Steady-state ($t\rightarrow\infty$) cavity photon number distribution under PReSPA, as probed by ancilla qubit spectroscopy.  In this measurement, PReSPA is applied nearly at all times, only  briefly interrupted by spectroscopy probes once every 2 ms.  The peak amplitudes confirm that odd photon number parity is permanently stabilized, and also reveals the presence of spurious excitation processes in the cavity.   
    \textbf{d}, Steady-state cavity photon number distribution when only the mixing comb of PReSPA is applied, as in (\textbf{b}). Under this configuration, a transmon $\gamma_\uparrow$ jump may add just one spurious photon in the cavity, leading to an effective cavity heating rate out of its vacuum state $\gamma_{01}\approx\gamma_\uparrow$ (when the transmon decay rate through the reservoir  $\gamma_m\gg 1/T_{1q}$).  The relative probability of $\ket{0}$ vs.~$\ket{1}$ informs the balance between the cavity decay rate $1/T_{1A}=1.9$ ms$^{-1}$ and the heating rate $\gamma_{01}$. \textbf{e}, Blue triangles: cavity heating rate $\gamma_{01}$, as measured with the technique described in (\textbf{d}), as a function of the Rabi amplitude of a single $\ket{0e0}\leftrightarrow\ket{1g1}$ mixing tone.  $\gamma_{01}$ initially follows the expected values corresponding to transmon $\gamma_{\uparrow}=1.4$ ms$^{-1}$ (green dashed curve), but additional heating effects due to the microwave pump is observed at higher $\Omega$.
    }
    \label{fig:transmon_heating}
\end{figure*}

\begin{figure*}[tbp]
    \centering
    \includegraphics[scale=.50]{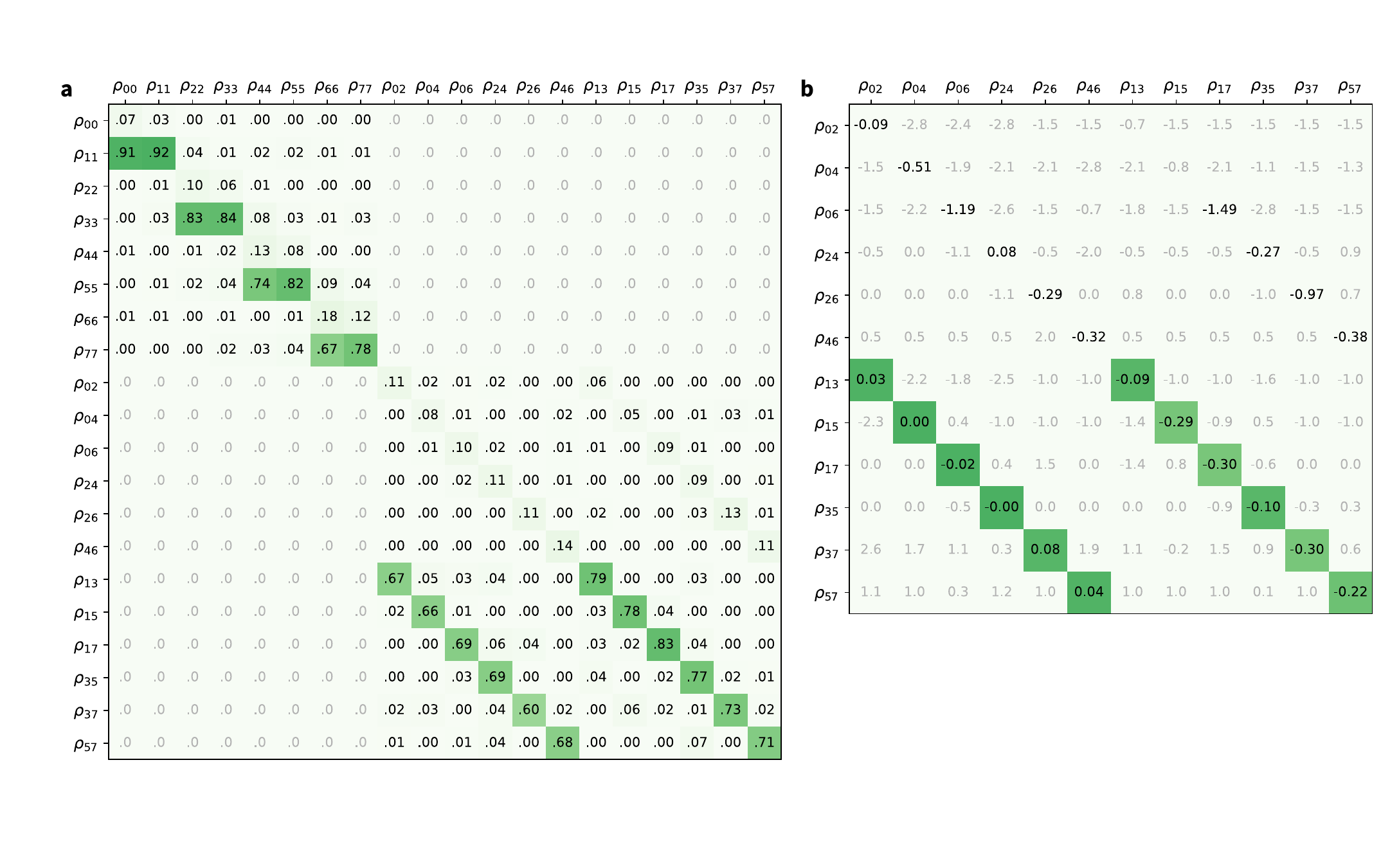}
    \caption{ \textbf{Process $\chi$ matrix block for $25$ $\mu s$ of PReSPA.} The matrix converts elements of input density matrices, top axis, to output density matrix elements, left axis, expressed in the Fock state basis.
    \textbf{a}, Amplitude values of the $\chi$ matrix elements. 
    The upper left block ($\chi_{nn,mm}$) describes the conversion of diagonal elements of the input and output density matrices, which is associated with transfer of photon occupation probabilities calculated from transmon spectroscopy experiments (as in Fig.~2d,e).  The lower right block ($\chi_{nm,kl}$) describes the conversion of relevant off-diagonal elements of density matrices, which is calculated from Wigner tomography and density matrix reconstruction~\cite{cahill_density_1969} (as in Fig.~3).  The greyed blocks are assumed to be zero due to the absence of interference between the four conversion paths in PReSPA. 
    \textbf{b}, Phase values of the $\chi$ matrix elements for the lower right block in \textbf{a}.  
    For best illustration of the PReSPA process, the phases are reported in the full rotating frame where all Fock states have zero energy. Values in grey are measured but not statistically significant as the corresponding amplitude value is not large enough.  In this frame, as prescribed by Eq.~(\ref{eq:PReSPA}), PReSPA requires zero phase for the six $\chi_{nm,(n+1)(m+1)}$ elements representing the coherence of the even-to-odd conversion process, which is accomplished by our PReSPA calibration.  The diagonal elements $\chi_{nm,nm}$ representing the preservation of odd-state superpositions should have zero phase by definition.  Their systematic deviation from zero was caused by parameter drift in the experiment as that block of data was acquired at a later time than the earlier rotating frame calibration.
    \label{fig:process_matrix}
    }
\end{figure*}

Since the simulation quantitatively captures all the error channels displayed is the experiment, we can evaluate the performance of our PReSPA pumping scheme when the system parameter improves.  With better cavity and transmon coherence, for the same static and driven Hamiltonian parameters, the logical error rates due to transmon decay and the second photon decay will be suppressed quadratically, as expected for an ideal first-order QEC protocol.  However, other error channels due to higher-order nonlinearity and drive-tone selectivity will not scale down equally.  Therefore, reduced $\chi_q$, $\kappa$, $\Omega$ and $\lambda$ should be chosen to partially trade away the gain from coherence improvement to suppress these error channels.
This is analogous to the situation in active QEC experiments, where less frequent parity measurements may be desirable when coherence time improves~\cite{ofek_extending_2016, hu_quantum_2019}.

Extended Data Fig.~\ref{fig:gain}a demonstrates the simulated AQEC performance including all the imperfections intrinsic to a transmon-based PReSPA scheme, which is computed using a specific set of Hamiltonian parameters (which corresponds to 50\% lower PReSPA rates and 80\% lower dispersive shift than the experiment).  The result shows logical lifetime 7-9 times longer than the physical photon loss time for a $\bar{n}=3.4$ T4C code, or 40-80\% above the breakeven point, using current state-of-the-art cavity and transmon of the same style as in our experiment.  Extrinsic effects such as thermal excitation of the transmon is not included, but in Extended Data Fig.~\ref{fig:gain}b we show for a specific attainable case of $T_{1A}=1$ ms, $T_{1q}=T_{1\phi}=100$ $\mu$s, QEC breakeven can be achieved with a few percent of transmon thermal populations.  

In Extended Fig.~\ref{fig:gain}a, assuming constant $\lambda$ and $\chi_q$, the QEC gain factor decreases at the longest $T_{1A}$ ($>2$ ms).  This is because of increased weight of error contribution from off-resonance excitation of the transmon due to imperfect frequency selectivity of the transmon comb.  Further reduction of $\lambda$ (accompanied with smaller reduction of $\chi_q$) is needed to further improve the gain factor.  A more promising alternative in this high-coherence regime is to use $\ket{2n+1,f,0}$ instead of $\ket{2n,e,0}$ as the first intermediate level of PReSPA, which can immediately reduce all frequency-selectivity-related errors by a factor of 4 at the expense of 2$\times$ faster transmon decay (a relatively small contribution in this regime). 


\begin{figure*}[tbp]
    \centering
    \includegraphics[scale=.50]{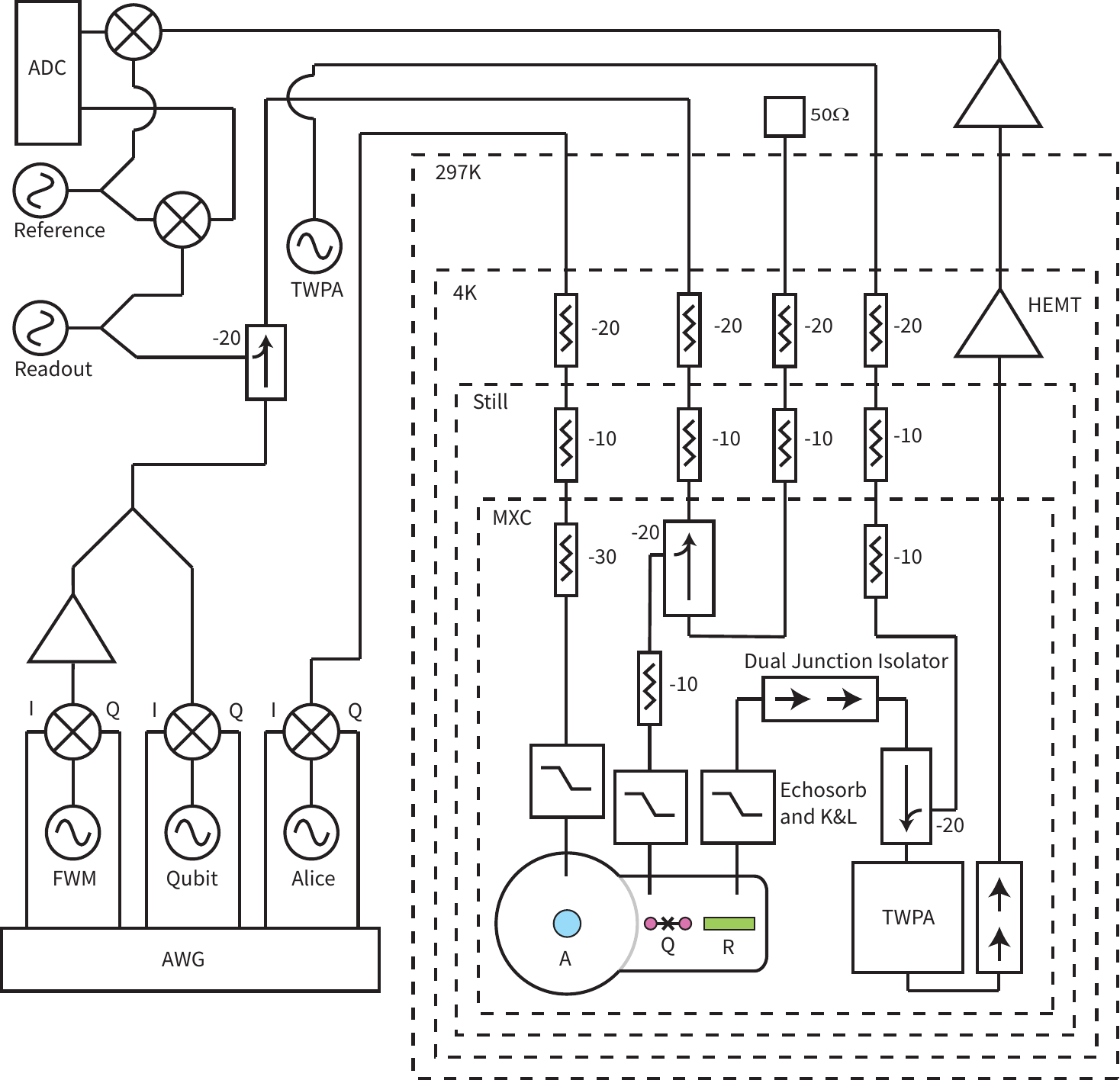}
    \caption{\textbf{Fridge Diagram.} Our experiments are performed in an Oxford Triton-500 dilution refrigerator.   Our device is mounted at the mixing chamber (MXC) stage of the refrigerator with a nominal temperature of about 10 mK.  We perform a heterodyne transmission measurement of the readout cavity to determine the transmon state. Coherent control signals for both cavity $A$ and the transmon are generated by IQ modulation. Each of the two frequency combs implementing PReSPA is generated by software addition of four frequencies playing on one set of IQ channels (the mixing comb has its own dedicated IQ channels while the transmon comb shares the same IQ channels as the regular transmon qubit gates). We use a first-stage Traveling Wave Parametric Amplifier~\cite{macklin_nearquantum-limited_2015} (TWPA) from  MIT Lincoln Laboratory, as well as a High-Electron-Mobility Transistor (HEMT) amplifier at the 4K stage. Our experiment is housed inside a MuMETAL shield to protect it from external magnetic fields. All lines connected to the experiment are filtered with a K\&L 10 GHz low-pass filter and an Eccosorb low-pass filter.
    \label{fig:fridge}
    }
\end{figure*}


\begin{thebibliography}{10}


\bibitem{lidar_quantum_2013}
Lidar, D.~A. \& Brun, T.~A. (eds.).
\newblock \emph{Quantum {Error} {Correction}},  (Cambridge University Press,
\vspace*{0mm}2013), 1st ed.

\bibitem{schindler_experimental_2011}
Schindler, P. et~al.
\newblock Experimental {Repetitive} {Quantum} {Error} {Correction}.
\newblock \emph{Science} \textbf{332}, 1059--1061 (2011).

\bibitem{cramer_repeated_2016}
Cramer, J. et~al.
\newblock Repeated quantum error correction on a continuously encoded qubit by
  real-time feedback.
\newblock \emph{Nature Communications} \textbf{7}, 11526 (2016).

\bibitem{kelly_state_2015}
Kelly, J. et~al.
\newblock State preservation by repetitive error detection in a superconducting
  quantum circuit.
\newblock \emph{Nature} \textbf{519}, 66--69 (2015).

\bibitem{ofek_extending_2016}
Ofek, N. et~al.
\newblock Extending the lifetime of a quantum bit with error correction in
  superconducting circuits.
\newblock \emph{Nature} \textbf{536}, 441--445 (2016).

\bibitem{hu_quantum_2019}
Hu, L. et~al.
\newblock Quantum error correction and universal gate set operation on a
  binomial bosonic logical qubit.
\newblock \emph{Nature Physics} \textbf{15}, 503--508 (2019).

\bibitem{andersen_repeated_2020}
Andersen, C.~K. et~al.
\newblock Repeated quantum error detection in a surface code.
\newblock \emph{Nature Physics} pp. 1--6 (2020).

\bibitem{ahn_continuous_2002}
Ahn, C., Doherty, A.~C. \& Landahl, A.~J.
\newblock Continuous quantum error correction via quantum feedback control.
\newblock \emph{Physical Review A} \textbf{65}, 042301 (2002).

\bibitem{atalaya_continuous_2020}
Atalaya, J. et~al.
\newblock Continuous quantum error correction for evolution under
  time-dependent {Hamiltonians}.
\newblock \emph{arXiv:}2003.11248 (2020).

\bibitem{kerckhoff_designing_2010}
Kerckhoff, J., Nurdin, H.~I., Pavlichin, D.~S. \& Mabuchi, H.
\newblock Designing {Quantum} {Memories} with {Embedded} {Control}: {Photonic}
  {Circuits} for {Autonomous} {Quantum} {Error} {Correction}.
\newblock \emph{Physical Review Letters} \textbf{105}, 040502 (2010).

\bibitem{kapit_hardware-efficient_2016}
Kapit, E.
\newblock Hardware-{Efficient} and {Fully} {Autonomous} {Quantum} {Error}
  {Correction} in {Superconducting} {Circuits}.
\newblock \emph{Physical Review Letters} \textbf{116}, 150501 (2016).

\bibitem{reiter_dissipative_2017}
Reiter, F., Sørensen, A.~S., Zoller, P. \& Muschik, C.~A.
\newblock Dissipative quantum error correction and application to quantum
  sensing with trapped ions.
\newblock \emph{Nature Communications} \textbf{8}, 1822 (2017).

\bibitem{albert_pair-cat_2019}
Albert, V.~V. et~al.
\newblock Pair-cat codes: autonomous error-correction with low-order
  nonlinearity.
\newblock \emph{Quantum Science and Technology} \textbf{4}, 035007 (2019).

\bibitem{sarovar_continuous_2005}
Sarovar, M. \& Milburn, G.~J.
\newblock Continuous quantum error correction by cooling.
\newblock \emph{Physical Review A} \textbf{72}, 012306 (2005).

\bibitem{brune_observing_1996}
Brune, M. et~al.
\newblock Observing the {Progressive} {Decoherence} of the ``{Meter}'' in a
  {Quantum} {Measurement}.
\newblock \emph{Physical Review Letters} \textbf{77}, 4887--4890 (1996).

\bibitem{mundhada_experimental_2019}
Mundhada, S. et~al.
\newblock Experimental {Implementation} of a {Raman}-{Assisted} {Eight}-{Wave}
  {Mixing} {Process}.
\newblock \emph{Physical Review Applied} \textbf{12}, 054051 (2019).

\bibitem{reinhold_error-corrected_2020}
Reinhold, P. et~al.
\newblock Error-corrected gates on an encoded qubit.
\newblock \emph{Nature Physics} pp. 1--5 (2020).

\bibitem{ma_error-transparent_2020}
Ma, Y. et~al.
\newblock Error-transparent operations on a logical qubit protected by quantum
  error correction.
\newblock \emph{Nature Physics} pp. 1--5 (2020).

\bibitem{knill_theory_1997}
Knill, E. \& Laflamme, R.
\newblock Theory of quantum error-correcting codes.
\newblock \emph{Physical Review A} \textbf{55}, 900--911 (1997).

\bibitem{lihm_implementation-independent_2018}
Lihm, J.-M., Noh, K. \& Fischer, U.~R.
\newblock Implementation-independent sufficient condition of the
  {Knill}-{Laflamme} type for the autonomous protection of logical qudits by
  strong engineered dissipation.
\newblock \emph{Physical Review A} \textbf{98}, 012317 (2018).

\bibitem{poyatos_quantum_1996}
Poyatos, J.~F., Cirac, J.~I. \& Zoller, P.
\newblock Quantum {Reservoir} {Engineering} with {Laser} {Cooled} {Trapped}
  {Ions}.
\newblock \emph{Physical Review Letters} \textbf{77}, 4728--4731 (1996).

\bibitem{krauter_entanglement_2011}
Krauter, H. et~al.
\newblock Entanglement {Generated} by {Dissipation} and {Steady} {State}
  {Entanglement} of {Two} {Macroscopic} {Objects}.
\newblock \emph{Physical Review Letters} \textbf{107}, 080503 (2011).

\bibitem{barreiro_open-system_2011}
Barreiro, J.~T. et~al.
\newblock An open-system quantum simulator with trapped ions.
\newblock \emph{Nature} \textbf{470}, 486--491 (2011).

\bibitem{kienzler_quantum_2015}
Kienzler, D. et~al.
\newblock Quantum harmonic oscillator state synthesis by reservoir engineering.
\newblock \emph{Science} \textbf{347}, 53--56 (2015).

\bibitem{leghtas_confining_2015}
Leghtas, Z. et~al.
\newblock Confining the state of light to a quantum manifold by engineered
  two-photon loss.
\newblock \emph{Science} \textbf{347}, 853--857 (2015).

\bibitem{lescanne_exponential_2020}
Lescanne, R. et~al.
\newblock Exponential suppression of bit-flips in a qubit encoded in an
  oscillator.
\newblock \emph{Nature Physics} \textbf{16}, 509-513 (2020).

\bibitem{koch_charge-insensitive_2007}
Koch, J. et~al.
\newblock Charge-insensitive qubit design derived from the {Cooper} pair box.
\newblock \emph{Physical Review A} \textbf{76}, 042319 (2007).

\bibitem{reagor_quantum_2016}
Reagor, M. et~al.
\newblock A quantum memory with near-millisecond coherence in circuit {QED}.
\newblock \emph{Physical Review B} \textbf{94}, 014506 (2016).

\bibitem{axline_architecture_2016}
Axline, C. et~al.
\newblock An architecture for integrating planar and {3D} {cQED} devices.
\newblock \emph{Applied Physics Letters} \textbf{109}, 042601 (2016).

\bibitem{chuang_bosonic_1997}
Chuang, I.~L., Leung, D.~W. \& Yamamoto, Y.
\newblock Bosonic quantum codes for amplitude damping.
\newblock \emph{Physical Review A} \textbf{56}, 1114--1125 (1997).

\bibitem{terhal_towards_2020}
Terhal, B.~M., Conrad, J. \& Vuillot, C.
\newblock Towards scalable bosonic quantum error correction.
\newblock \emph{Quantum Science and Technology} \textbf{5}, 043001 (2020).

\bibitem{mirrahimi_dynamically_2014}
Mirrahimi, M. et~al.
\newblock Dynamically protected cat-qubits: a new paradigm for universal
  quantum computation.
\newblock \emph{New Journal of Physics} \textbf{16}, 045014 (2014).

\bibitem{heeres_implementing_2017}
Heeres, R.~W. et~al.
\newblock Implementing a universal gate set on a logical qubit encoded in an
  oscillator.
\newblock \emph{Nature Communications} \textbf{8}, 1--7 (2017).

\bibitem{cohen_degeneracy-preserving_2017}
Cohen, J., Smith, W.~C., Devoret, M.~H. \& Mirrahimi, M.
\newblock Degeneracy-{Preserving} {Quantum} {Nondemolition} {Measurement} of
  {Parity}-{Type} {Observables} for {Cat} {Qubits}.
\newblock \emph{Physical Review Letters} \textbf{119}, 060503 (2017).

\bibitem{pinotsi_single_2008}
Pinotsi, D. \& Imamoglu, A.
\newblock Single {Photon} {Absorption} by a {Single} {Quantum} {Emitter}.
\newblock \emph{Physical Review Letters} \textbf{100}, 093603 (2008).

\bibitem{vlastakis_deterministically_2013}
Vlastakis, B. et~al.
\newblock Deterministically {Encoding} {Quantum} {Information} {Using}
  100-{Photon} {Schrödinger} {Cat} {States}.
\newblock \emph{Science} \textbf{342}, 607--610 (2013).

\bibitem{schuster_resolving_2007}
Schuster, D.~I. et~al.
\newblock Resolving photon number states in a superconducting circuit.
\newblock \emph{Nature} \textbf{445}, 515--518 (2007).

\bibitem{serniak_direct_2019}
Serniak, K. et~al.
\newblock Direct {Dispersive} {Monitoring} of {Charge} {Parity} in
  {Offset}-{Charge}-{Sensitive} {Transmons}.
\newblock \emph{Physical Review Applied} \textbf{12}, 014052 (2019).

\bibitem{peterer_coherence_2015}
Peterer, M.~J. et~al.
\newblock Coherence and {Decay} of {Higher} {Energy} {Levels} of a
  {Superconducting} {Transmon} {Qubit}.
\newblock \emph{Physical Review Letters} \textbf{114}, 010501 (2015).

\bibitem{grimm_stabilization_2020}
Grimm, A. et~al.
\newblock Stabilization and operation of a {Kerr}-cat qubit.
\newblock \emph{Nature} \textbf{584}, 205--209 (2020).

\bibitem{ma_path-independent_2020}
Ma, W.-L. et~al.
\newblock Path-{Independent} {Quantum} {Gates} with {Noisy} {Ancilla}.
\newblock \emph{Physical Review Letters} \textbf{125}, 110503 (2020).

\bibitem{puri_stabilized_2019}
Puri, S. et~al.
\newblock Stabilized {Cat} in a {Driven} {Nonlinear} {Cavity}: {A}
  {Fault}-{Tolerant} {Error} {Syndrome} {Detector}.
\newblock \emph{Physical Review X} \textbf{9}, 041009 (2019).

\bibitem{doucot_physical_2012}
Douçot, B. \& Ioffe, L.~B.
\newblock Physical implementation of protected qubits.
\newblock \emph{Reports on Progress in Physics} \textbf{75}, 072001 (2012).

\bibitem{gyenis_experimental_2019}
Gyenis, A. et~al.
\newblock Experimental realization of an intrinsically error-protected
  superconducting qubit.
\newblock \emph{arXiv:}1910.07542 (2019).

\bibitem{macklin_nearquantum-limited_2015}
Macklin, C. et~al.
\newblock A near–quantum-limited {Josephson} traveling-wave parametric
  amplifier.
\newblock \emph{Science} \textbf{350}, 307--310 (2015).

\bibitem{grimsmo_quantum_2020}
Grimsmo, A.~L., Combes, J. \& Baragiola, B.~Q.
\newblock Quantum {Computing} with {Rotation}-{Symmetric} {Bosonic} {Codes}.
\newblock \emph{Physical Review X} \textbf{10}, 011058 (2020).

\bibitem{leghtas_hardware-efficient_2013}
Leghtas, Z. et~al.
\newblock Hardware-{Efficient} {Autonomous} {Quantum} {Memory} {Protection}.
\newblock \emph{Physical Review Letters} \textbf{111}, 120501 (2013).

\bibitem{susskind_quantum_1964}
Susskind, L. \& Glogower, J.
\newblock Quantum mechanical phase and time operator.
\newblock \emph{Physics Physique Fizika} \textbf{1}, 49--61 (1964).

\bibitem{michael_new_2016}
Michael, M.~H. et~al.
\newblock New {Class} of {Quantum} {Error}-{Correcting} {Codes} for a {Bosonic}
  {Mode}.
\newblock \emph{Physical Review X} \textbf{6}, 031006 (2016).

\bibitem{wang_schrodinger_2016}
Wang, C. et~al.
\newblock A {Schrödinger} cat living in two boxes.
\newblock \emph{Science} \textbf{352}, 1087--1091 (2016).

\bibitem{nigg_black-box_2012}
Nigg, S.~E. et~al.
\newblock Black-{Box} {Superconducting} {Circuit} {Quantization}.
\newblock \emph{Physical Review Letters} \textbf{108}, 240502 (2012).

\bibitem{sun_tracking_2014}
Sun, L. et~al.
\newblock Tracking photon jumps with repeated quantum non-demolition parity
  measurements.
\newblock \emph{Nature} \textbf{511}, 444--448 (2014).

\bibitem{werschnik_quantum_2007}
Werschnik, J. \& Gross, E. K.~U.
\newblock Quantum optimal control theory.
\newblock \emph{Journal of Physics B: Atomic, Molecular and Optical Physics}
  \textbf{40}, R175--R211 (2007).

\bibitem{glaser_training_2015}
Glaser, S.~J. et~al.
\newblock Training {Schrödinger}’s cat: quantum optimal control.
\newblock \emph{The European Physical Journal D} \textbf{69}, 279 (2015).

\bibitem{gollub_monotonic_2008}
Gollub, C., Kowalewski, M. \& de~Vivie-Riedle, R.
\newblock Monotonic {Convergent} {Optimal} {Control} {Theory} with {Strict}
  {Limitations} on the {Spectrum} of {Optimized} {Laser} {Fields}.
\newblock \emph{Physical Review Letters} \textbf{101}, 073002 (2008).

\bibitem{khaneja_optimal_2005}
Khaneja, N., Reiss, T., Kehlet, C., Schulte-Herbrüggen, T. \& Glaser, S.~J.
\newblock Optimal control of coupled spin dynamics: design of {NMR} pulse
  sequences by gradient ascent algorithms.
\newblock \emph{Journal of Magnetic Resonance} \textbf{172}, 296--305 (2005).

\bibitem{kirchmair_observation_2013}
Kirchmair, G. et~al.
\newblock Observation of quantum state collapse and revival due to the
  single-photon {Kerr} effect.
\newblock \emph{Nature} \textbf{495}, 205--209 (2013).

\bibitem{cahill_density_1969}
Cahill, K.~E. \& Glauber, R.~J.
\newblock Density {Operators} and {Quasiprobability} {Distributions}.
\newblock \emph{Physical Review} \textbf{177}, 1882--1902 (1969).

\end{thebibliography}


\end{document}


\title{Supplementary Materials for \\
Protecting a Bosonic Qubit with Autonomous Quantum Error Correction}
\author{Jeffrey M.~Gertler}
\affiliation{Department of Physics, University of Massachusetts-Amherst, Amherst, MA, 01003, USA}
\author{Brian Baker}
\affiliation{Department of Physics and Astronomy, Northwestern University, Evanston, IL, 60201, USA}
\author{Juliang Li}
\affiliation{Department of Physics, University of Massachusetts-Amherst, Amherst, MA, 01003, USA}
\author{Shruti Shirol}
\affiliation{Department of Physics, University of Massachusetts-Amherst, Amherst, MA, 01003, USA}
\author{Jens Koch}
\affiliation{Department of Physics and Astronomy, Northwestern University, Evanston, IL, 60201, USA}
\author{Chen Wang*}
\affiliation{Department of Physics, University of Massachusetts-Amherst, Amherst, MA, 01003, USA}
\email{wangc@umass.edu}

\date{\today}


\maketitle

\section{Bosonic Code for Approximate AQEC}
\subsection{Applying PReSPA to Schr\"odinger-cat code ($|\alpha|\gg1$)}

Odd-parity cat states of the form $\ket*{C^-_{\alpha}} = (\ket{\alpha} - \ket{-\alpha})/N$, $\ket*{C^-_{i\alpha}} = (\ket{i\alpha} - \ket{-i\alpha})/N$
with large photon number, $|\alpha|^2\gg1$, can be used as cat-codes \cite{leghtas_hardware-efficient_2013, mirrahimi_dynamically_2014} that allow for perfect retrieval of information with an appropriate unitary. As shown in the following, single-photon loss followed by ideal PReSPA extended to infinite photon numbers
\begin{equation}\label{PReSPA_infinity}
\hat{\Pi}'_{eo}=\sum_{n=0}^{\infty}\ket{2n+1}\bra{2n}    
\end{equation}
preserves the capability to recover the initially stored quantum information. 



Consider a general initial state $\ket{\psi(0)} = x\ket*{C^-_{\alpha}} + y\ket*{C^-_{i\alpha}}$ in the cat encoding. Within a given time period $t$, this state undergoes stochastic evolution governed by photon-loss events and continuous AQEC driving, resulting in a piece-wise deterministic process for $\ket{\psi(t)}$. Jump events are caused by photon loss, instantaneously followed by a PReSPA operation. In our analysis, we neglect the short time span $t_e \ll T_{1A}$ involved in the PReSPA operation itself. Under these conditions, the compound effect of photon loss and PReSPA is described by the jump operator $\hat{\Pi}'_{eo}\hat{a} = \sum_{n=0}^{\infty}\sqrt{2n+1}\ket{2n+1}\!\bra{2n+1}$. The deterministic time-evolution between jumps, as viewed from the rotating frame, is governed by the non-unitary operator
\begin{equation}\label{determ_evolve}
    \hat{V}(t) = \text{exp}\Big(-\frac{t}{2T_{1A}} \hat{a}^\dag\hat{a}\Big). 
\end{equation}
Note that $\hat{V}(t)$ decreases the cat size, 
$\alpha \to \tilde{\alpha}(t) = \alpha e^{-t/2T_{1A}}$. By contrast, photon loss/PReSPA events cause an increase in the mean photon number. For $j$ loss/recovery events during time $t$, the final state is given by
\begin{equation}\label{catpsij}
    \ket{\psi_j(t)} = x\ket*{C^-_{\tilde{\alpha},j}} + y\ket*{C^-_{i\tilde{\alpha},j}},
\end{equation}
where $\ket*{C^-_{\tilde{\alpha},j}} = (\hat{\Pi}'_{eo}\hat{a})^j\ket{C^-_{\tilde{\alpha}}}/N$. Although the latter are not perfect cat states anymore, they maintain parity and near orthogonality.
Upon averaging over trajectories with different jump numbers $j$ , one finds for the density matrix of the joint transmon-cavity system:
\begin{equation}\label{catcode_mixedstate}
    \hat{\rho}(t) = \ket{g}\!\bra{g}\otimes\sum_j p_j(t)\ket{\psi_j(t)}\!\bra{\psi_j(t)} = \ket{g}\!\bra{g}\otimes(x + ye^{i\pi a^\dag a /2})\sum_j p_j(t)\ket*{C^-_{\tilde{\alpha},j}}\!\bra*{C^-_{\tilde{\alpha},j}}(x^* + y^*e^{-i\pi a^\dag a /2}).
\end{equation}
Here, $p_j(t)$ is the probability for $j$ loss/recovery events to occur during time $t$.

For recovery of the amplitudes $x$ and $y$ to succeed, $\ket*{C^-_{\tilde{\alpha},j}}$ and $\ket*{C^-_{i\tilde{\alpha},k}}$ should be orthogonal $\forall j,k$. Since $\bra*{C^-_{i\tilde{\alpha},j}}\ket*{C^-_{\tilde{\alpha},k}} \propto e^{-|\tilde{\alpha}|^2}\sin(|\tilde{\alpha}|^2)$, (near-)orthogonality is given for $|\alpha|^2\gg1$. Accordingly, the cavity Hilbert space splits into two orthogonal subspaces, with $\ket*{C^-_{\tilde{\alpha},j}} \in \mathcal{H}_1$, $\ket*{C^-_{i\tilde{\alpha},j}} \in \mathcal{H}_2$, and the prescriptions 
$\ket{g}\otimes\ket*{C^-_{\tilde{\alpha},j}} \rightarrow \ket{g}\otimes\ket*{C^-_{\tilde{\alpha},j}}$,  $\ket{g}\otimes\ket*{C^-_{i\tilde{\alpha},j}} \rightarrow \ket{e}\otimes e^{-i\pi a^\dag a /2}\ket*{C^-_{i\tilde{\alpha},j}}$, are compatible with a unitary decoding transformation. Once applied to the state \eqref{catcode_mixedstate}, this yields
\begin{equation}\label{decoded_cat}
    \hat{\rho}_d = \hat{U}_d\hat{\rho}\hat{U}^\dag_d = (x\ket{g} + y\ket{e})(x^*\bra{g} + y^*\bra{e})\otimes \sum_j p_j\ket*{C^-_{\tilde{\alpha},j}}\!\bra*{C^-_{\tilde{\alpha},j}},
\end{equation}
leaving the transmon qubit in the desired pure state $x\ket{g} + y\ket{e}$.

\subsection{Evolution of truncated 4-component cat (T4C) states under photon loss and PReSPA}

Our T4C encoding is an instance of the single-cavity binomial code~\cite{michael_new_2016}.  It can also be understood as a cat code with $|\alpha|^2\approx3.5$ truncated at a photon number cutoff at 7.  Here, 
we encode the quantum information in the superposition $\ket{\psi(0)} = x\ket{0_L} + y\ket{1_L}$, where the logical code words are
\begin{align}\label{code0}
    \ket{0_L} = C_1\ket{1} + C_5\ket{5}, \qquad 
    \ket{1_L} = C_3\ket{3} + C_7\ket{7}.
\end{align}
The effective jump operator for loss/recovery events is a truncation of the original $\hat{\Pi}'_{eo}\hat{a}$, namely
$\hat{\Pi}_{eo}\hat{a} = \ket{1}\!\bra{1} + \sqrt{3}\ket{3}\!\bra{3} + \sqrt{5}\ket{5}\!\bra{5} + \sqrt{7}\ket{7}\!\bra{7}$.
The resulting quantum trajectory given $j$ events is
\begin{equation}\label{psij}
    \ket{\psi_j(t)} = \frac{1}{N_j(t)}\Big[x\sum_{n=1,5} + y\sum_{n=3,7} \Big] C_n n^{j/2}e^{-nt/2T_{1A}}\ket{n},
\end{equation}
where $N_j(t)$ is the time-dependent normalization factor. Note that the trajectory \eqref{psij} exhibits no dependence on the specific jump times, because the jump operator $\hat{\Pi}_{eo}\hat{a}$ commutes with the deterministic evolution operator \eqref{determ_evolve}. 
To compare Eq.\ \eqref{psij} with \eqref{catpsij}, we rewrite the trajectory state as a superposition of two states from the orthogonal subspaces $\mathcal{H}_{15}=\mathrm{span} \{\ket{1}, \ket{5}\}$ and  $\mathcal{H}_{37}=\mathrm{span} \{\ket{3}, \ket{7}\}$, namely
\begin{equation}\label{final_state}
    \ket{\psi_j(t)} = x\, n^{15}_j(t)\ket*{\psi_j^{15}(t)} + y\, n^{37}_j(t)\ket*{\psi_j^{37}(t)}.      
\end{equation}
Here, $\ket{\psi_j^{kl}(t)} = \sum_{n=k,l}C_nn^{j/2}e^{-n t/2T_{1A}}\ket{n}/[n_j^{kl}N_j(t)]$ are orthonormal, and 
%
$n^{kl}_j(t) = \| \sum_{n=k,l}C_n n^{j/2}e^{-nt/2T_{1A}}\ket{n}\| /N_j(t)$. 

As a side note, although the mean photon numbers $\bar{n}$ varies from trajectory to trajectory, the ensemble-averaged photon number, $\langle\bar{n}\rangle$, of a cavity state is conserved under the combined effect of $\hat{\Pi}_{eo}\hat{a}$.  This is because the operator $\hat{\Pi}_{eo}\hat{a}$ is a (diagonal) dephasing-type of jump operator that does not extract or supply energy to the system.  One can also conveniently prove this by checking the diagonal vector of the Lindblad master equation, and see for any density matrix $\rho$: $\text{diag}[\dot\rho]\propto\text{diag}[\mathcal{L}(\hat{\Pi}_{eo}\hat{a}, \rho)]=0$.

\subsection{T4C decoding unitary for approximate quantum information retrieval}
Given the distortion due the $n^{kl}_j(t)$  coefficients, as well as the rotations within the $\mathcal{H}_{15}$ and $\mathcal{H}_{37}$ subspaces, to what degree does it remain possible to decode the original quantum information? The answer and mitigation strategy depends on the elapsed time $t$. Here, we focus on short times $t/T_{1A} \ll1$, for which only small jump numbers $j$ are relevant. [For $t/T_{1A}\ll1$, $p_j(t)$ is peaked at $j=0$ and falls off rapidly as $j$ increases. These jump probabilities are given by $p_j(t) = (t/T_{1A})^jN^2_j(t)/j!$]


We divide our discussion into two parts, addressing the distortion coefficients and spurious rotations separately. Starting with the former, we note that the distortion via $n^{kl}_j(t)$ is not observed in Eq.\ \eqref{catpsij} for the cat code. For the T4C scheme the distortion originates from the truncation to small photon numbers. We may mitigate this issue by choosing the code-word amplitudes $C_n$ is such a way as to minimize distortion, i.e., make the relevant $n_j^{kl}(t)$ as close to unity as possible. 
Expanding numerator and denominator of $(n_0^{kl})^2$ \big[$(n_0^{15})^2$ or $(n_0^{37})^2$\big] for $t/T_{1A}\ll1$ yields the intermediate expression
%
%
%
%
\begin{equation}
[n^{kl}_0(t)]^2 = \frac{1 - t/T_{1A} \langle n \rangle_{kl} + \frac{(t/T_{1A})^2}{2}\langle n^2 \rangle_{kl} + \mathcal{O}((t/T_{1A})^3)}{x^2\big[1 -t/T_{1A} \langle n \rangle_{15} + \frac{(t/T_{1A})^2}{2}\langle n^2 \rangle_{15} + \mathcal{O}((t/T_{1A})^3)\big] + y^2\big[1 -t/T_{1A} \langle n \rangle_{37} + \frac{(t/T_{1A})^2}{2}\langle n^2 \rangle_{37} + \mathcal{O}((t/T_{1A})^3)\big]},    
\end{equation}    
where $\langle n^m \rangle_{kl}$ is the $m$-th moment of the photon number probability distribution, with probabilities $|C_k|^2$ and $|C_l|^2 = 1-|C_k|^2$. This expression reveals that $n_0^{kl} = 1$ up to second order in $t/T_{1A}$ if the mean and second moments obey $\langle n \rangle_{15} = \langle n \rangle_{37}$ and $\langle n^2 \rangle_{15} = \langle n^2 \rangle_{37}$. These two conditions, along with the normalization requirement, fully determine the optimal choices for the code-word amplitudes: $C_1 = C_7 = 1/2$ and $C_3 = C_5 = \sqrt{3}/2$. (The choice $C_1 = \sqrt{0.35}$, $C_5 = \sqrt{0.65}$, $C_3 = \sqrt{0.9}$, and $C_7 = \sqrt{0.1}$ used in the experiments yields mean photon numbers of 3.6 and 3.4, and second moments of 16.6 and 13, constituting a reasonable, though not optimum, choice.) Expansion to higher orders in $t/T_{1A}$ as well as consideration of jump numbers $j>0$ lead to additional conditions $\langle n^m \rangle_{15}=\langle n^m \rangle_{37}$ with $m>2$, which cannot be satisfied anymore.

Second, we consider the spurious rotations within each code-word subspace, while focusing on the construction of a suitable decoding unitary. For $t/T_{1A}\ll 1$ and small jump numbers, this unitary $\hat{U}_d$ should approximately map the state \eqref{final_state} to a product state, in which the transmon occupies the state $x\ket{g} + y\ket{e}$, i.e. 
\begin{equation}\label{joint_space_trajectory}
\ket{\Psi_j(t)} = \ket{g}\otimes\big(xn^{15}_j(t)\ket{\psi^{15}_j(t)} + yn^{37}_j(t)\ket{\psi^{37}_j(t)}\big) \;\xrightarrow{\hat{U}_d} \;\ket{\Psi_j(t)}_d \approx \big(x\ket{g} + y\ket{e}\big)\otimes\ket{\phi_j},     
\end{equation}
where the cavity states $\ket{\phi_j(t)}$ must be consistent with the unitarity of $U_d$, but can otherwise be chosen at will. 
To approximate this transformation, we select orthonormal bases of the two code-word subspaces,  $\mathcal{H}_{15}=\mathrm{span} \{\ket{u_0}, \ket{u_1}\}$ and  $\mathcal{H}_{37}=\mathrm{span} \{\ket{v_0}, \ket{v_1}\}$, and choose $U_d$ to map
\begin{equation}
\ket{g}\otimes\ket{u_0} \rightarrow \ket{g0}, \quad
\ket{g}\otimes\ket{u_1} \rightarrow \ket{g1}, \quad
\ket{g}\otimes\ket{v_0} \rightarrow \ket{e0}, \quad
\ket{g}\otimes\ket{v_1} \rightarrow \ket{e1}.    
\end{equation}
We only specify $\hat{U}_d$ for these 4 basis states relevant to the T4C code, leaving the rest undetermined. Decomposed in this basis, the evolved T4C state reads
\begin{equation}\label{trajectory}
    \ket{\Psi_j(t)} = \ket{g}\otimes x\,n^{15}_j(t)\big(\cos\theta_j\ket{u_0}  
    + \sin\theta_j\ket{u_1}) + y\,n^{37}_j(t)\big(\cos\varphi_j\ket{v_0} + \sin\varphi_j\ket{v_1}\big),
\end{equation}
where the angles $\theta_j$ and $\varphi_j$ parametrize the states $\ket*{\psi_j^{kl}(t)}$ (see main text, Extended Data Fig.\ 1a). The state resulting after application of the decoding unitary is
\begin{equation}
    \ket{\Psi_j(t)}_d = \hat{U}_d\ket{\Psi_j(t)} =  x\,n^{15}_j(t)\ket{g}\otimes\big(\cos\theta_j\ket{0} + \sin\theta_j\ket{1}\big)   
    + y\,n^{37}_j(t)\ket{e}\otimes\big(\cos\varphi_j\ket{0} + \sin\varphi_j\ket{1}\big).
\end{equation}
The decoding fidelity can in principle be optimized by a suitable choice of the cavity states $\ket{u_k}$, $\ket{v_k}$. Here, we omit optimization and simply select the original code-words as basis states, $\ket{u_0} = \ket{0_L}$ and $\ket{v_0} = \ket{1_L}$. This is appropriate for $t/T_{1A}\ll1$ where small jump numbers dominate. From series expansion of $\cos^2\theta_0 = |\bra{0_L}\ket{\psi_0^{15}}|^2$ and $\cos^2\varphi_0 = |\bra{1_L}\ket{\psi_0^{37}}|^2$,  one finds $\theta_0 = \varphi_0 + \mathcal{O}((t/T_{1A})^2)$ if the optimum code-word amplitudes $C_n$ are employed. Imperfections in decoding do arise in the small-time limit for larger jump numbers.
Following the decoding operation, the reduced density matrix of the transmon takes on the form
\begin{equation}
    \hat{\rho}_q = \text{Tr}_{\text{cavity}}\left(\hat{U}_d\sum_jp_j\ket{\Psi_j}\!\bra{\Psi_j}\hat{U}^\dag_d\right) = 
    \sum_jp_j\begin{pmatrix}
    |x|^2(n^{15}_j)^2 & x^*yn^{15}_jn^{37}_j\cos(\theta_j-\varphi_j) \\ xy^*n^{15}_jn^{37}_j\cos(\theta_j-\varphi_j) & |y|^2(n^{37}_j)^2
    \end{pmatrix},
\end{equation}
where time dependence is implied but not shown explicitly to simplify notation.

$\hat{\rho}_q$ is a good short-time approximation to the intended state $\ket{\psi_q}\!\bra{\psi_q}$. The deviation between the two is quantified by the process fidelity that is obtained from averaging the state-transfer fidelity $\mathcal{F}_{xy}(t) = |\langle \psi_q | \hat{\rho}_q(t) | \psi_q \rangle |^2$ over the six cardinal points on the Bloch sphere and then rescaled to the range from 1/4 to 1,  
\begin{equation}
\mathcal{F}_{\text{process}}(t) =  \frac{3}{2}\langle \mathcal{F}_{xy}(t)\rangle - \frac{1}{2},
\end{equation}
where $\langle \mathcal{F}_{xy}(t)\rangle$ is the Bloch sphere average.
%
Future work may include optimization of this fidelity with respect to the choice of decoding basis states.

\section{Optimal control for unitary operations and state transfer}
Pulses for state preparation and decoding are constructed using quantum optimal control (QOC) \cite{werschnik_quantum_2007,glaser_training_2015, gollub_monotonic_2008,heeres_implementing_2017}, designed to steer the time evolution of a quantum system in order to maximize the fidelity of the desired state transfer or unitary operation. More specifically, QOC  iteratively adjusts the set of control pulses $\{u_1(t),\ldots,u_M(t)\}$ in a way that minimizes the deviations between the realized unitary $\hat{U}$ and the target unitary $\hat{U}_T$ (or between realized state $\ket{\psi}$ and target state $\ket{\psi_T}$). The concrete quantity to be minimized is the cost functional $C[\{u_k(t)\}]$, which incorporates the process or state infidelity, as well as additional cost contributions crucial for achieving realistic pulse shapes. We employ an automatic-differentiation extension\cite{leung_speedup_2017} of GRAPE (Gradient Ascent Pulse Engineering) \cite{khaneja_optimal_2005}, which performs minimization via gradient descent. In the following, we provide additional details on the implementation.

For numerics, the control fields are reduced to a finite number of optimization parameters by discretization of the time axis: the pulse duration $t_d$ is split into a number $N\gg1$ of small time intervals (here: $\delta t = 1$\,ns). The corresponding discrete time instances are denoted $t_n = n\delta t$, along with the control amplitudes evaluated at these times, $u_{kn}=u_k(t_n)$. The latter can now be grouped into a vector $\mathbf{u}=(u_{kn}) \in \mathbb{R}^{MN}$ containing all optimization parameters. The primary quantity for optimization is either the process fidelity (target unitary, such as for decoding) or the state-transfer fidelity (target state, such as for preparation of specific cavity state). These are defined as 
\begin{align}
    \mathcal{F}_{\text{decoding}}(\mathbf{u}) &= \frac{1}{D^2}|\text{Tr}(\hat{U}^\dag_T\hat{U}(\mathbf{u}))|^2, \\ 
    \mathcal{F}_{\text{prep}}(\mathbf{u}) &= |\bra{\psi_T}\ket{\psi(\mathbf{u})}|^2,
\end{align}
where $D = 7$, the dimension of the subspace considered for the decoding operation. The realized unitary or state is computed by decomposing the evolution operator $\hat{U}_f$ governing the full duration $t_d$ into short-time propagators $\hat{U}_n = \text{exp}(-i\hat{H}_n\delta t)$ describing the unitary evolution of the system from time $t_n$ to $t_n + \delta t$, 
\begin{equation}
    \hat{U}_f = \hat{U}_N \hat{U}_{N-1} \cdots \hat{U}_1 \hat{U}_0.
\end{equation}
 Each short-time propagator only involves the Hamiltonian evaluated at time $t_n$, 
\begin{equation}
    \hat{H}_n = \hat{H}(t=t_n) = \hat{\mathcal{H}}_0 + \sum_{k=1}^M u_{kn}\hat{\mathcal{H}}_{k}
\end{equation} 
with $\hat{\mathcal{H}}_0$ denoting the time-independent system Hamiltonian and  $\{\hat{\mathcal{H}}_{k}\}$ the set of control operators. In our case, the latter correspond to the transmon and cavity operators $\{\hat{\sigma}_x, \hat{\sigma}_y, \hat{x}_A, \hat{p}_A\}$, and the static system Hamiltonian is given by
\begin{equation}
\hat{\mathcal{H}}_0 = - \frac{\chi_q}{2} \hat{a}^\dagger\hat{a}\hat{\sigma}_z - \frac{K}{2}\hat{a}^\dagger\hat{a}^\dagger\hat{a}\hat{a} - \frac{\chi'_q}{4}\hat{a}^\dagger\hat{a}^\dagger\hat{a}\hat{a}\hat{\sigma}_z   
\end{equation}
in the appropriate rotating frame. The coefficients $K$, $\chi_q$, $\chi'_q$ parametrize the self-Kerr, $4^{\text{th}}$-order cross-Kerr, and $6^{\text{th}}$-order cross-Kerr terms, respectively. 



In addition to the primary contribution $C_1=1-\mathcal{F}$ to the cost function, i.e.\ the process or transfer infidelity, it is beneficial to incorporate secondary cost contributions which favor pulse shapes compatible with experimental capabilities and constraints. One secondary cost contribution,
\begin{equation}
    C_2 = \sum_{k, n} | u_{kn} - u_{k n-1} |^2,
\end{equation}
penalizes rapid temporal changes of the pulse envelopes, thus helping to limit overall pulse bandwidth. Another secondary cost contribution we employ is 
\begin{equation}
    C_3 = \sum_{k,n} |u_{kn}|^2,
\end{equation}
which limits the overall pulse power. Finally, we introduce a cost for occupation of certain higher states $\ket{\psi_F}$ (``forbidden" states). This can, for example, minimize leakage of the transmon into states outside the $\{\ket{g}, \ket{e}\}$ logical subspace, and generally help avoid any population close to the Hilbert space truncation level (truncation levels used were 3 and 24 for the transmon and cavity Hilbert space, respectively). The corresponding  cost contribution is
\begin{equation}
    C_4 = \sum_{n} |\bra{\psi_F}\ket{\psi_n}|^2,
\end{equation}
where $\ket{\psi_n}$ is the state at time $t_n$. The total cost function to be minimized is a weighted sum of the individual contributions,
\begin{equation}
C(\mathbf{u}) = C_1(\mathbf{u}) + \alpha_2C_2(\mathbf{u}) + \alpha_3C_3(\mathbf{u}) + \alpha_4C_4(\mathbf{u}).
\end{equation}

In its basic form, gradient descent minimizes $C(\mathbf{u})$ by repeatedly updating $\mathbf{u}$ with a vector opposite to the cost function gradient,
\begin{equation}\label{gradient_descent_eq}
     \mathbf{u}_{p+1} =  \mathbf{u}_p - \eta_p\frac{\partial C(\mathbf{u}_p)}{\partial \mathbf{u}_p}.
\end{equation} 
Here, we use ADAM \cite{kingma_adam_2017}, a ``momentum accelerated" gradient-descent method which updates $\mathbf{u}$ using a moving average of both the gradient direction and magnitude. To make convergence properties more stable, we apply the common strategy of a decaying learning rate $\eta_p = \eta_0e^{-\beta p}$. Gradients $\partial C(\mathbf{u}_p)/\partial \mathbf{u}_p$ are computed using a TensorFlow-supported automatic differentiation algorithm \cite{baydin_automatic_2017}. Rather than hard-coding analytical gradients, this algorithm decomposes $C(\mathbf{u}_p)$ into a computational graph of elementary functions, whose known derivatives are extracted via the chain rule through a back-propagation process.

For bootstrapping, we used Gaussian white noise as initial pulses $\mathbf{u}_0$ in order to avoid bias towards any particular pulse shape. Computations were carried out on a CPU with a step count of $N = t_d / \delta t$. For state preparation, $t_d = 1$\,$\mu$s and for decoding, $t_d = 2$\,$\mu$s. The total number of iterations to achieve sufficient fidelity convergence typically ranged from 1,000-2,000, thus requiring only moderate computation times of a few hours.

\section{Error sources of the corrected logical qubit}
The errors in the autonomously corrected logical qubit can be divided into two main categories: 1) errors unrelated to single photon loss that are not yet corrected by our AQEC scheme, and 2) errors caused by various failure modes of our AQEC scheme while it corrects single photon loss.  A breakdown of the contributions from various error mechanisms is listed in Extended Data Table III.  
The total effect is consistent with the experimentally measured longitudinal ($366\pm10$ $\mu$s) and transverse ($256\pm7$ $\mu$s) relaxation times of the logical qubit within 5\%. 

\subsection{Photon-loss unrelated errors}

Cavity dephasing is not correctable with the PReSPA scheme.  The dominant source of cavity dephasing is spontaneous excitation (the rate $\gamma_\uparrow$) of the ancilla transmon from its ground state to excited states, which we believe is caused by stray radiation from the microwave transmission line. 
In addition to its dephasing effect to the cavity (due to the dispersive Hamiltonian $-\chi_q \hat{a}^\dagger\hat{a}\frac{\sigma_z}{2}$), an ancilla $\gamma_\uparrow$ event may 
trigger an odd-to-even photon addition process under PReSPA due to the relative lack of photon-number selectivity in the second stage of the pumping process (Extended Data Fig.~7a).  This process erroneously switches the cavity to the even-parity sub-space, which will immediately gain another photon under PReSPA.  Effectively, each $\gamma_\uparrow$ jump has a substantial probability to induce the addition of two photons to the cavity state, moving the cavity state between the $\{\ket{1}, \ket{5}\}$ subspace and the $\{\ket{3}, \ket{7}\}$ subspace, which is a bit flip error for logical pole states.  From competition of relevant rates in this multi-stage process, we estimate this probability to be about 50\%. 
For logical equator states, $\gamma_\uparrow$ produces a transverse relaxation error by fully dephasing the cavity.  In total, our AQEC scheme, as currently constructed, leaves the logical qubit modestly more susceptible to $\gamma_\uparrow$ of the ancilla compared with a bare cavity using Fock states $\ket{0}$ and $\ket{1}$ to store information.

Strong parametric pumping as used in our mixing tone increases spurious transmon excitation.  We observe an increase in transmon $\gamma_\uparrow$ rate by about 30\% under long-time pumping of the mixing comb, from 1.4 ms$^{-1}$ in the idle state to 1.8 ms$^{-1}$, as extracted from a set of specifically designed experiments probing spurious cavity excitations (Extended Data Fig.~7b-e).  We note that all mixing tones in PReSPA combined (computed as the sum of $|\xi_i|$) are still weaker than reported thresholds for causing spurious transmon excitations in prior experiments~\cite{gao_programmable_2018,rosenblum_cnot_2018, touzard_coherent_2018}, so it is likely that the observed heating effect is extrinsic and reducible by better thermalization.  In addition, we estimate that the pump-induced linear displacement to the transmon operator $\hat{q}$ should contribute negligibly to spurious transmon excitations.  However, we cannot rule out contributions from non-perturbative dynamics of a driven Josephson circuit~\cite{lescanne_escape_2019}, especially considering a comb of multiple pump tones may cause additional complexities in the dynamics beyond what has been reported so far.  This scenario corresponds to a fundamental limitation of using a transmon as a parametric mixer, and motivates the adoption of novel types of ancillas in future AQEC experiments.  Nevertheless, a future iteration of better-optimized transmon-based PReSPA protocol with better coherence properties will employ weaker (rather than stronger) pumping than the current demonstration, hence alleviating this concern of pump-induced spurious excitations. (See Methods-AQEC Simulations.)

Separately, the transmon comb of PReSPA has an intrinsic effect of exciting the transmon off-resonantly when the cavity is in the odd-parity code space.  With our parameters, this effect is weak compared with other $\gamma_\uparrow$ effects and difficult to experimentally characterize, but we determine it to be slightly lower than 0.2 ms$^{-1}$ from numerical simulations.  
This excitation rate can be estimated as $\gamma_m(\lambda/\chi_q)^2$, where $\gamma_m$ is the transmon decay rate through the four-wave mixing process.  In the long run, this intrinsic effect requires trade-off in the QEC operation speed.  It is important to note that its scaling is to the third power of the QEC operation speed if we scale $\lambda$, $\Omega$, $\kappa$ proportionally, as we confirmed in simulations.

\subsection{Photon-loss correction efficiency}

PReSPA is designed to correct for single photon loss, which occurs on a timescale of $T_{1A}/\bar{n}\approx150$ $\mu$s in our T4C-encoded logical qubit and induces both longitudinal and transverse errors.  An intuitive quantity to evaluate the efficacy of this correction process is the QEC success rate, $S_l$ or $S_t$, as we defined in the main text.  It represents the fraction of photon-loss errors whose impact to the logical qubit is successfully eliminated by the AQEC scheme.  For the convenience of arguments, we imagine that each photon loss occurring to the T4C qubit, when uncorrected, induce a single depolarization error from anywhere on the Bloch sphere (equivalent to a longitudinal relaxation error for pole state and a transverse relaxation error for an equator state).  This is in fact a decent approximation for the uncorrected T4C code.  

In the absence of any error unrelated to photon loss, the corrected logical qubit lifetime is equal to the inversely averaged relaxation time of the 6 cardinal-point states of the Bloch sphere:
\begin{align}
    \tau_C = \frac{6}{4\Gamma_t+2\Gamma_l} =\frac{T_{1A}}{\bar{n}}\frac{1}{1-\frac{2}{3}S_t-\frac{1}{3}S_l}
\end{align}
where we have used $\Gamma_i=\frac{\bar{n}}{T_{1A}}(1-S_i)$ by our definition of $S_i$.  On the other hand, the Fock $\ket{0/1}$ encoding has lifetime of $\tau_{01}=\frac{3}{2}T_{1A}$.  Hence to reach break-even the overhead factor due to redundant encoding one has to overcome is $3\bar{n}/2$.  From $\tau_C>\tau_{01}$, a necessary requirement for QEC break-even is:
\begin{align}
    \frac{1}{3}S_l+\frac{2}{3}S_t > 1-\frac{2}{3}\bar{n}^{-1}
    \label{eq:threshold}
\end{align}
We note that due to intrinsic off-resonant excitation of transmon in our current PReSPA scheme, error channels not attributed to photon loss cannot be fully eliminated.  This effect, together with other types of uncorrectable errors, means that a greater success rate than indicated by Eq.~(\ref{eq:threshold}) is needed to achieve break-even.  However, this minimum threshold still serves as a useful benchmark to consider the performance of a QEC process in correcting the errors it is designed to correct.

An ideal (infinitely-fast) implementation of PReSPA dissipation using a perfect ancilla can fully eliminate longitudinal errors and correct transverse errors at 97.5\% success probability (see Methods-Approximate AQEC).  
Experimentally, there are multiple other failure modes associated with the photon-loss-correction process, as listed in Extended Data Table III, adding up to an estimated total of 11\% and 24\% failure rates for correcting pole and equator states in this work.  {(This corresponds to $S_l=89\%$ and $S_t=76\%$ as discussed in the main text.)}

Relaxation of the ancilla and the loss of a second photon during the correction process account for a significant part of the failure rates.  During each error-correction process, the probability of transmon $T_{1}$ decay can be computed as $\int P_e(t)dt/T_{1q}=7\%$, where $P_e(t)$ is the transmon excitation probability (Extended Data Fig.~3c-f).  The probability of a second photon decay can be estimated as $t_{eo}\bar{n}/T_{1A}=5.6\%$, where $t_{eo}$ is the halftime of the error-correction process.  Since a second photon decay is a bit-flip error, its reduction to $S_l$ is twice of its occurrence rate, or about 11\%.  These failure rates are dictated by the competition between the speed of the QEC operation and the hardware coherence, and can be expected to improve proportionally with better transmon and cavity lifetime.

In addition, there are several failure modes associated with the competing time scales in PReSPA.
Firstly, the pumping process can be activated virtually by a neighboring mixing tone that is $2\chi_q$ detuned from the resonance condition, causing cavity dephasing.  This occurs at an estimated 3\% probability based on the relative rates of the virtual (unintended) \textit{vs}.~resonant (intended) four-wave-mixing transitions, and scales as $(\kappa/2\chi_q)^2$. 
In our experiment, $\kappa$ can be reduced by at least $\sim$20\% without compromising the speed of the AQEC operation, hence improving overall AQEC performance.

Secondly, 
due to the stochastic nature of the dissipative correction process, the cavity spends an uncertain period of time in the even-parity error space.  This time uncertainty $\delta t_{eo}$ is smaller than but scales linearly with the average correction time $t_{eo}$.  It multiplies with the self-Kerr ($K$) of the cavity Hamiltonian to produce a phase uncertainty to the PReSPA operator, and hence a logical dephasing error to the QEC process.

Furthermore, 
in the presence of other higher-order nonlinear effects from $\chi'_q$ and $\chi_{Ar}$, the two groups of intermediate states in the PReSPA dynamics ($\ket{2n,e,0}$ and $\ket{2n+1,g,1}$) are not equally spaced.  One might expect them to cause which-path information leakage due to reservoir photon emissions at different frequencies.  However, the actual effect is more subdued:  In a dissipative $\Lambda$ transition, the emission frequency is solely dictated by the energy difference of initial/final states and energy supplied by the drives, not the intermediate levels.  (The energy differences due to Kerr is already accounted for, see the paragraph above).  This is a technique to promote path-independence using slightly-virtual transitions~\cite{wang_autonomous_2019}.  In this scenario, because the quantum states do not transit through the intermediate states exactly on resonance, there is a subtle geometric phase uncertainty causing a relatively small dephasing effect.  The scaling of this residual error has been analyzed in detail in Appendix C of Ref.~\cite{wang_autonomous_2019}, which applied to the present cases is $(\chi'_q/\gamma_m)^2$ and $(\chi_{Ar}/\kappa)^2$ respectively.  Here $\gamma_m\approx0.35$ $\mu$s$^{-1}$ is the rate of transmon decay via the reservoir due to the mixing comb.

Our AQEC process is insensitive to dephasing ($T_\phi$) error of the ancilla.  Owing to the mirrored dynamics of the four photon-addition paths in PReSPA, a phase jump of the transmon ancilla during the photon addition resets the process but does not cause cavity dephasing.  The net effect is a small reduction of the correction rate of PReSPA, which can be compensated by choosing a larger $\lambda$.  Similarly, low-frequency (\textit{i.e.~}$1/f$) noise of the ancilla does not affect PReSPA performance to the first order 
either.  This insensitivity to ancilla dephasing has an interesting correspondence to the scenario in measurement-based bosonic QEC, where a $T_{\phi}$ error leads to a parity (error syndrome) measurement error but does not propagate forward to the cavity state~\cite{ofek_extending_2016}.  This type of error in discrete parity measurements in principle can be  mitigated by repetition and majority voting, which however, amplifies other types of errors (such as ancilla $T_1$) and are typically undesirable in experiments~\cite{ofek_extending_2016,hu_quantum_2019}.  In comparison, our continuous AQEC protocol effectively repeats its error detection process automatically until it succeeds with minimal overhead.

\section{Error Recovery Test}
In the process of tuning up PReSPA, we developed an experiment to evaluate the efficiency of PReSPA in performing a single QEC operation. To do this we prepare with an OCT pulse the ``ideal error state":
\begin{align}
    \ket{\psi_{e}} = \frac{1}{\sqrt{2}}(C_1\ket{0} 
     + e^{i\phi_3} C_3\ket{2} 
    + e^{i\phi_5} C_5 \ket{4} 
    + e^{i\phi_7} C_7 \ket{6}).
\end{align}
This is a state that the ideal PReSPA operator (Eq.~(2) of the main text) will correct back to the $\ket{X}_L$ state after 25 $\mu$s. The phases $\phi_3$, $\phi_5$ and $\phi_7$ are chosen to compensate for the phase accumulation due to the cavity self-Kerr.  

After applying 25 $\mu s$ of PReSPA, we use the decoding unitary to map the logical state to the transmon, and measure the $\hat{\sigma}_x$ projection. To compensate for the loss in fidelity due to encoding and decoding we perform a separate experiment of encoding and immediately decoding a logical $\ket{x}_L$ state before taking a $\hat{\sigma}_x$ projection.  By comparing the results of these two measurements we measure a normalized fidelity of the recovered state $F = 67\pm2\%$.  

This measurement is analogous to characterization of the PReSPA process $\chi$-matrix, which includes additional errors accumulated over the 25 $\mu$s process time in addition to the one-time parity restoration process.  It is therefore in direct agreement with the 67\% average preservation of phase coherence in Fig.~3.  It is also consistent with the success probability $S_t=76\%$ for a one-time PReSPA operation after accounting for the compounded 10\% fidelity loss from all transverse error channels.

\section{PReSPA optimizer for logical equator states}  

To achieve a better choice of PReSPA control parameters (the amplitudes and phases of microwave tones) we implement an empirical optimization routine. The ultimate goal of our optimization is to decrease the decay rate of the process fidelity to increase AQEC efficiency. For this we initialize a logical state, apply 144 $\mu$s of PReSPA, decode the state and perform transmon tomography. We can optimize this fidelity by slightly varying a PReSPA control parameter and performing this procedure. By iterating over each PReSPA control parameter, we can gradually optimize the AQEC scheme. Experimentally we find that the logical equator states ($\ket{X}_L$, $\ket{-X}_L$, $\ket{Y}_L$, $\ket{-Y}_L$) are more susceptible to small changes in PReSPA  parameters than the pole states ($\ket{Z}_L$, $\ket{-Z}_L$). Because of this, we choose our optimization cost function to be a mean of the fidelities of these four states.

This optimization was implemented to choose control parameters used for PReSPA-2 in Extended Data Fig.~5. We achieved a slight gain in logical coherence time over PReSPA-1, whose control parameters were only chosen using the calibration experiments described in the methods section. While better for logical lifetimes, we observed that the PReSPA-2 rates ($\Omega_n$ and $\lambda_n$) were less well matched than those for PReSPA-1. Further work is needed to understand how a dissipative operator that deviates from Eq.~(2) can achieve better AQEC performance.\\

\section{All About Readout}
To readout the transmon state we perform a heterodyne measurement on the readout cavity state looking for a disspersive shift in its frequency dependant on transmon state. This signal is amplified by a TWPA, HEMT and room temperature amplifier but we don't have the signal to noise ratio to confidently perform single shot readout. Instead we measure the transmission signal with the transmon prepared in the nominal ground and excited states to measure the contrast by which we convert raw, averaged signals to excited state population measurements.

The nominal ground state of the transmon is its equilibrium prepared by waiting for a sufficiently long time ($>10 T_{1q}$) for the transmon to decay, and the nominal excited state is obtained by $\pi$-pulsing the qubit from its nominal ground state.  Since our transmon has a 5\% excited state population in equilibrium, this means that all reported quantum state fidelity should be understood as scaled up by 10\% compared to the true pure target state.  However, since all the main results reported in this manuscript focus on the relative change between two states, such as process fidelity or decay time, this factor affects measurements of both states proportionally and therefore does not affect the quantitative conclusions of our experiment.  For example, the relaxation rates measured in Fig 4b are independent of the way this error is handled.   

Due to the cross-Kerr between our readout mode and cavity A we see a slight photon-number dependent reduction in readout contrast. Usually this error happens on measurements where we are selectively measuring the transmon frequency to probe cavity $A$ photon number. By removing the selective transmon rotation and performing readout we can measure the shift in readout contrast due to the cavity occupation.

This cross-Kerr readout error also shows up in Wigner parity measurements. For these we perform two measurements, one mapping $\ket{\text{even}}_A\rightarrow\ket{g}_q$ and $\ket{\text{odd}}_A\rightarrow\ket{e}_q$ while the other performs the map $\ket{\text{even}}_A\rightarrow\ket{e}_q$ and $\ket{\text{odd}}_A\rightarrow\ket{g}_q$. With both of these measurements we can calculate the impact of cavity occupation on readout contrast.

In Fig. 2f we are measuring at times with significant transmon excitation probabilities. This means that the measurement contrast will be effected.  We apply the same background measurement that calculates cross-Kerr readout shifts to measure transmon excitation probability. With this we can accurately show the photon addition probability over time. This transmon excitation data is also useful to accurately calculate $\lambda$ and $\Omega$ as shown in Extended Data Fig. 3.

Because our OCT state preparation pulses have errors we perform a correction to some measurements based on this state preparation error. Fig. 2f, Extended Data Fig. 3 and 8 have numbers re-normalized to these state preparation errors. As these figures are characterizing the PReSPA process we didn't want to include these inaccuracies. These state preparation errors are not compensated for in any figure discussing AQEC such as Fig. 4d.


\begin{figure*}[tbp]
    \centering
    \includegraphics[scale=0]{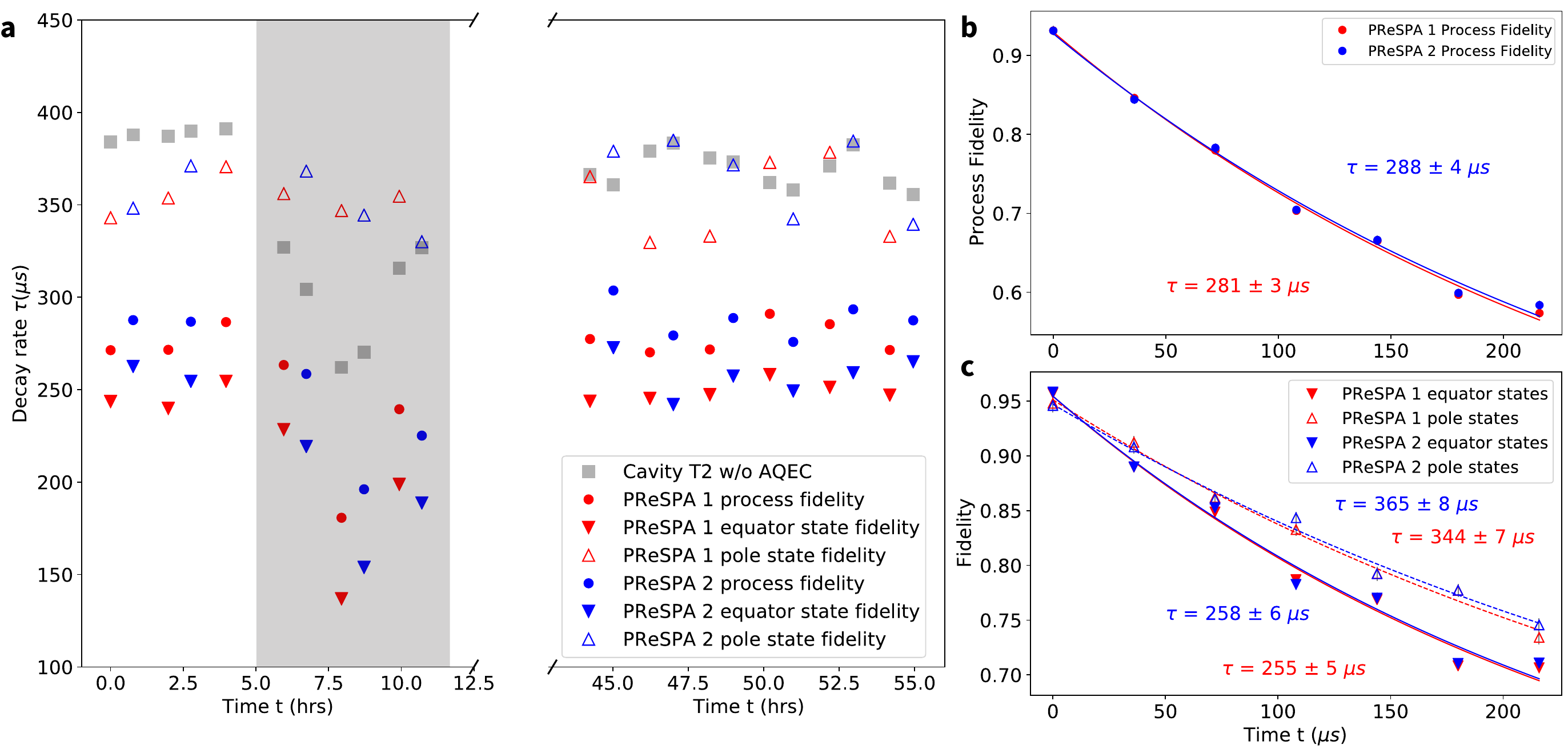}
\end{figure*}
